\documentclass[aps,pre,reprint,amsmath,amssymb,superscriptaddress,longbibliography]{revtex4-2}

\usepackage{graphicx}
\usepackage{bm}
\usepackage{xcolor}
\usepackage{tikz}
\usetikzlibrary{arrows.meta,positioning,decorations.pathmorphing,calc}
\usepackage[colorlinks=true,linkcolor=blue!60!black,citecolor=blue!60!black,urlcolor=blue!60!black]{hyperref}

\newcommand{\dd}{\mathrm{d}}
\newcommand{\ee}{\mathrm{e}}
\newcommand{\Dx}{\mathcal{D}_x}            
\newcommand{\DD}{\mathcal{D}_D}            
\newcommand{\rs}{\rho^{s}}                 
\newcommand{\Jx}{J_x^{s}}
\newcommand{\JD}{J_D^{s}}
\newcommand{\Path}{\Gamma}                 
\newcommand{\T}{\mathcal{T}}               
\newcommand{\Smf}{\Delta S_{\mathrm{m\&f}}}
\newcommand{\Smed}{\Delta S_{\mathrm{med}}}
\newcommand{\Stot}{\Delta S_{\mathrm{tot}}}
\newcommand{\Ssys}{\Delta S_{\mathrm{sys}}}
\newcommand{\avg}[1]{\left\langle #1\right\rangle}
\newcommand{\strat}{\circ}
\newcommand{\OM}{\mathbf{S}}           
\newcommand{\Di}{\mathcal{D}_i}        

\begin{document}

\title{Entropy production decomposition and fluctuation theorems under continuous feedback: From prescribed to stochastic protocols}

\author{Long Him Cheung}
\email{lhcheung@umd.edu}
\affiliation{Department of Physics, University of Maryland, College Park, Maryland 20742, USA}
\affiliation{Institute for Physical Science and Technology, University of Maryland, College Park, Maryland 20742, USA}

\date{\today}

\begin{abstract}
When a system is driven by a prescribed protocol, the entropy production can be decomposed into the housekeeping and non-autonomous contributions, each satisfying a fluctuation theorem. Under feedback control, however, the feedback parameter depends on measurement outcomes rather than following a prescribed protocol. To study how the decomposition changes, we consider an overdamped Brownian particle at position $x$ with a low-pass-filtered measurement signal $D$ as the feedback parameter. Regarding each realization of $D$ as a stochastic protocol, we derive fluctuation theorems for the entropy production associated with the dynamics of $x$ and of $D$, and interpret them in terms of the information flow. For a solvable harmonic model, we map the dynamics onto an autonomous system in which the measurement and filtering process acts as an effective thermal bath. Feedback cooling thus corresponds to heat conduction from the physical bath through the particle to this colder bath, and the information flow can be interpreted as the entropy associated with the heat conduction between $x$ and $D$. We verify these results by numerical simulations.
\end{abstract}

\maketitle

\section{Introduction}
\label{sec:intro}
Stochastic thermodynamics is an extension of classical thermodynamics to small systems whose dynamics are significantly influenced by thermal fluctuations~\cite{Sekimoto2010,Seifert2012}. It has also been applied to systems involving measurement and feedback, in which the information obtained from measurements can be exploited to implement feedback for desired tasks, such as extracting work or cooling a stochastic system~\cite{Parrondo2015,Toyabe2010,Li2011,Gieseler2012}. In this framework, known as information thermodynamics~\cite{Parrondo2015}, the role of information is quantified by information-theoretic quantities, such as the mutual information~\cite{SagawaUeda2010,SagawaUeda2012}, the transfer entropy~\cite{Schreiber2000, ItoSagawa2013}, and the information flow~\cite{Allahverdyan2009,HorowitzEsposito2014}. Second-law-like inequalities and fluctuation theorems that incorporate these quantities have been established, generalizing the second law of thermodynamics to systems under measurement and feedback~\cite{SagawaUeda2010,SagawaUeda2012,ItoSagawa2013,HorowitzEsposito2014}.

For a feedback-controlled system, the feedback parameter plays the same role as the control parameter of a driven system. If the control parameter in a driven system is varied according to a prescribed protocol, the entropy production can be decomposed into the housekeeping and non-autonomous contributions, each of which satisfies a fluctuation theorem~\cite{HatanoSasa2001,SpeckSeifert2005,EspositoVdB2010PRL}. In a feedback-controlled system, however, the protocol is in general not prescribed. Instead, the feedback parameter evolves according to measurement outcomes. It is therefore natural to ask how this decomposition and the corresponding fluctuation theorems change when the protocol is determined by the measurement and feedback process.

In this paper, we investigate this question by considering an overdamped Brownian particle under continuous measurement and feedback. An observable $O(x)$ of the particle is measured continuously with Gaussian error; the measurement outcome is passed through a low-pass filter, and the filtered signal $D$ is used as the feedback parameter. Consequently, as shown in Sec.~\ref{sec:model}, the joint dynamics of the particle position $x$ and the filtered signal $D$ can be described as a bipartite Markov process~\cite{HorowitzEsposito2014,Hartich2014}, which allows the decomposition of the total entropy production into contributions associated with the dynamics of $x$ and of $D$. By regarding each realization of $D$ as a stochastic protocol, we derive fluctuation theorems for both contributions. These fluctuation theorems coincide with the stationary case of those for the partial entropy production derived in Ref.~\cite{RosinbergHorowitz2016}, where the reference dynamics was constructed to obtain these fluctuation theorems. Here, the same reference dynamics instead follows from extending the derivation for a prescribed protocol to a stochastic one, which allows a direct comparison between the entropy production decompositions for stochastic and prescribed protocols. In particular, the contribution associated with $D$ is the analog of the non-autonomous entropy production, whereas the contribution associated with $x$, unlike the housekeeping entropy production, does not vanish in the absence of non-conservative forces. Both contributions can also be expressed in terms of the information flow between the subsystems, which recovers the second-law-like inequalities bounded by the information flow~\cite{HorowitzEsposito2014,RosinbergHorowitz2016}.

We then apply these results to a solvable model of a free particle with harmonic feedback, for which the explicit expressions for all relevant quantities can be computed. We show that this model can be mapped exactly onto an autonomous system of two degrees of freedom coupled by a spring, each in contact with its own thermal bath. Under this mapping, the measurement and filtering process acts as an effective thermal bath, whose temperature and friction coefficient are determined by the measurement noise, the filter time constant, and the feedback stiffness. In this picture, feedback cooling corresponds to the effective bath being colder than the physical bath, and the information flow can be interpreted as the entropy associated with the heat conduction between the two subsystems. We verify the analytical results and the fluctuation theorems by numerical simulations.

\section{Model}
\label{sec:model}

We consider a one-dimensional overdamped Brownian particle in a potential $U(x)$, in contact with a thermal environment at temperature $T$ and with friction coefficient $\gamma$. Setting the Boltzmann constant $k_B=1$, the equation of motion is
\begin{equation}
  \dot x = -\frac{1}{\gamma}\partial_x U(x) + \sqrt{\frac{2T}{\gamma}}\,\xi_x(t),
  \label{eq:bare}
\end{equation}
where $\xi_x(t)$ is Gaussian white noise, and the noise amplitude $\sqrt{2T/\gamma}$ follows from the Einstein relation~\cite{Einstein1905,Kubo1966}.

Suppose that an observable $O(x)$ of the particle is measured continuously, and that the measurement is subject to Gaussian noise. This noise may represent instrumental error, or it may be added deliberately. We denote the measurement outcome $z(t)$, which is given by
\begin{equation}
  z(t) = O(x_t) + \sigma\,\xi_D(t),
  \label{eq:measurement}
\end{equation}
where $\sigma$ is the amplitude of the measurement noise and $\xi_D(t)$ is Gaussian white noise independent of $\xi_x(t)$. The two noises therefore satisfy
\begin{equation}
  \avg{\xi_i(t)}=0 \,\,,\,\, \avg{\xi_i(t)\,\xi_j(s)}=\delta_{ij}\,\delta(t-s) \,\,,\,\, i,j\in\{x,D\}.
  \label{eq:noise}
\end{equation}

\begin{figure}[!hbtp]
\centering
\begin{tikzpicture}[font=\small,>={Stealth[length=2mm]},
  box/.style={draw,rounded corners=2pt,minimum height=6mm,inner sep=3pt,align=center},
  bath/.style={draw,fill=gray!12,minimum width=12mm,minimum height=11mm,align=center,inner sep=2pt}]
  \node at (-0.55,1.25) {(a)};
  \node[box] (P) at (0.6,0.6) {particle $x$\\[-1pt]\scriptsize bath $(\gamma,T)$};
  \node[box] (M) at (3.25,0.6) {sensor\\[-1pt]\scriptsize $z=O(x)+\sigma\xi_D$};
  \node[box] (F) at (6.05,0.6) {low-pass filter\\[-1pt]\scriptsize $\tau_f$};
  \node[box] (V) at (3.25,-0.55) {feedback potential $V(x,D)$};
  \draw[->] (P) -- node[above,font=\scriptsize]{$O(x)$} (M);
  \draw[->] (M) -- node[above,font=\scriptsize]{$z$} (F);
  \draw[->] (F.south) |- node[pos=0.25,right,font=\scriptsize]{$D$} (V.east);
  \draw[->] (V.west) -| (P.south);
  \node at (-0.55,-1.45) {(b)};
  \node[bath] (B1) at (0.2,-2.25) {bath\\ $\gamma,\,T$};
  \node[circle,draw,fill=white,minimum size=6mm,inner sep=0] (x) at (1.75,-2.25) {$x$};
  \node[circle,draw,fill=white,minimum size=6mm,inner sep=0] (D) at (4.75,-2.25) {$D$};
  \node[bath] (B2) at (6.45,-2.25) {bath\\ $\gamma_D,\,T_D$};
  \draw (B1.east) -- (x.west);
  \draw (D.east) -- (B2.west);
  \draw[decorate,decoration={coil,aspect=0.6,segment length=2.2mm,amplitude=1.6mm}] (x.east) -- node[above=2.5mm,font=\scriptsize]{$V=\tfrac{k}{2}(x-D)^2$} (D.west);
  \node[font=\scriptsize,align=center] at (3.25,-3.2) {$\gamma_D=k\tau_f,\quad T_D=k\sigma^2/(2\tau_f)$};
\end{tikzpicture}
\caption{(a) Continuous measurement and feedback: the observable $O(x)$ is measured with white noise of strength $\sigma$, the signal is low-pass filtered with time constant $\tau_f$, and the filtered signal $D$ sets the feedback potential $V(x,D)$. (b) For $U=0$, $O(x)=x$, and $V=k(x-D)^2/2$ (Sec.~\ref{sec:harmonic}), the closed-loop dynamics is identical to that of two overdamped degrees of freedom coupled by a spring and attached to heat baths at temperatures $T$ and $T_D$.}
\label{fig:schematic}
\end{figure}
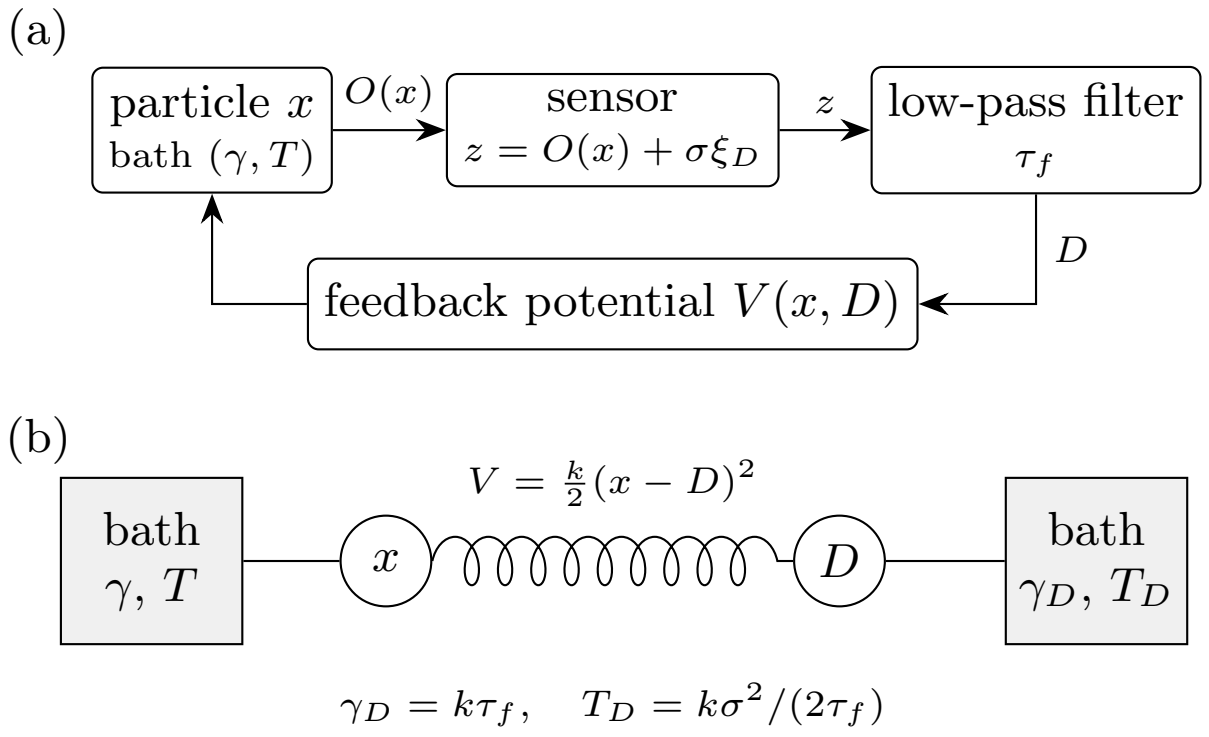

 Since $z(t)$ contains white noise, using it directly as a feedback parameter is in general not useful. Instead, we pass the measurement outcome through a low-pass filter with time constant $\tau_f$, i.e., bandwidth $1/\tau_f$, to smooth out the signal~\cite{AstromMurray2008}. The filtered signal $D(t)$ is therefore given by
\begin{equation}
  D(t) = D(0)\,\ee^{-t/\tau_f} + \frac{1}{\tau_f}\int_0^t \ee^{-(t-s)/\tau_f}\, z(s)\,\dd s .
  \label{eq:filter}
\end{equation}
Differentiating Eq.~\eqref{eq:filter} and substituting Eq.~\eqref{eq:measurement} gives a Langevin equation that governs the evolution of the filtered signal,
\begin{equation}
  \dot D = \frac{1}{\tau_f}\bigl[O(x)-D\bigr] + \frac{\sigma}{\tau_f}\,\xi_D(t) .
  \label{eq:Ddot}
\end{equation}
We then use the filtered signal as the feedback parameter to perform feedback by applying an additional potential $V(x,D)$ to the particle. The equations of motion for the joint dynamics of $(x,D)$ become
\begin{subequations}
\label{eq:EOM}
\begin{align}
  \dot x &= \frac{1}{\gamma}F(x,D) + \sqrt{\frac{2T}{\gamma}}\,\xi_x(t), \label{eq:EOMx}\\
  \dot D &= \frac{1}{\tau_f}\bigl[O(x)-D\bigr] + \frac{\sigma}{\tau_f}\,\xi_D(t), \label{eq:EOMD}
\end{align}
\end{subequations}
with $F(x,D) = -\partial_x\bigl[U(x)+V(x,D)\bigr]$. The joint dynamics of $(x,D)$ is therefore a two-dimensional Markov process with independent noises acting on $x$ and $D$, which is known as a bipartite process~\cite{HorowitzEsposito2014,Hartich2014}. A schematic is shown in Fig.~\ref{fig:schematic}(a).

For later convenience, we define the drift terms 
 \begin{equation}
     \mu_x = \frac{F}{\gamma} \quad,\quad \mu_D =  \frac{O(x)-D}{\tau_f} , \label{eq:mobility}
\end{equation}
and the diffusion coefficients
\begin{equation}
  \Dx = \frac{T}{\gamma} \quad,\quad \DD = \frac{\sigma^2}{2\tau_f^2} .
  \label{eq:diffusion}
\end{equation}
Here, $\DD$ is an effective diffusion coefficient determined by the measurement noise and the filter time constant. The joint probability density $\rho(x,D,t)$ obeys the Fokker-Planck equation
\begin{equation}
  \partial_t \rho = -\partial_x J_x - \partial_D J_D,
  \label{eq:FP}
\end{equation}
where the probability currents are given by
\begin{equation}
  J_x = \mu_x\rho - \Dx\,\partial_x\rho \quad,\quad
  J_D = \mu_D\rho - \DD\,\partial_D\rho .
  \label{eq:currents}
\end{equation}
A Fokker-Planck equation of the same form can also be derived from continuous measurement theory~\cite{Bhattacharyya2024}.

Unless stated otherwise, we assume that the joint dynamics admits a stationary distribution $\rs(x,D)$, and the stationary probability currents $J_x^s(x,D)$ and $J_D^s(x,D)$ satisfy
\begin{equation}
    \partial_t \rs = -\partial_x J_x^s - \partial_D J_D^s = 0.
\end{equation}
We also introduce the stationary potential $\phi(x,D) = -\ln\rs(x,D)$, which can be interpreted as the stochastic Shannon entropy of the joint system in the stationary state~\cite{Seifert2005}.

\section{Total entropy production}
\label{sec:total}

\subsection{Path probabilities}
\label{sec:total-path}

We use the path-integral representation of overdamped Langevin dynamics~\cite{OnsagerMachlup1953,Chernyak2006} with the Stratonovich (midpoint) discretization; see Appendix~\ref{app:path}. Products with a velocity are understood in the Stratonovich convention and are denoted by ``$\strat$''. For the joint dynamics defined by Eqs.~\eqref{eq:EOM}, the Onsager-Machlup function is
\begin{align}
  \mathbf{S}(x,\dot x,D,\dot D) ={}& \frac{\gamma}{4T}\Bigl(\dot x - \frac{F}{\gamma}\Bigr)^2 + \frac{1}{2\gamma}\partial_x F \nonumber\\
  &+ \frac{\tau_f^2}{2\sigma^2}\Bigl(\dot D - \frac{O(x)-D}{\tau_f}\Bigr)^2 - \frac{1}{2\tau_f} .
  \label{eq:OM}
\end{align}
Let $\Path=\{(x_t,D_t)\,|\,t\in[0,\T]\}$ denote a trajectory of $(x,D)$ over the time interval $[0,\T]$, and let $\Path^\dagger=\{(x^\dagger_t,D^\dagger_t)\,|\,t\in[0,\T]\}$ denote the corresponding time-reversed trajectory, with $(x^\dagger_t,D^\dagger_t)=(x_{\T-t},D_{\T-t})$. Since $D$ is a filtered signal that tracks the observable $O(x)$, we treat $D$ as even under time reversal. Assuming that the initial condition is sampled from $\rs$, the path probability is
\begin{equation}
  P[\Path] = \mathcal{N}\exp\Bigl(-\int_0^\T \mathbf{S}(x_t,\dot x_t,D_t,\dot D_t)\,\dd t\Bigr)\,\rs(x_0,D_0) ,
  \label{eq:pathprob}
\end{equation}
where $\mathcal{N}$ is a path-independent normalization factor.

Since the joint dynamics, Eq.~\eqref{eq:EOM}, contains no explicitly time-dependent control parameter, the reverse process is governed by the same equations of motion as the forward process. With the initial condition of the reverse process also sampled from $\rs$, the probability of observing the time-reversed trajectory is
\begin{equation}
  P[\Path^\dagger] = \mathcal{N}\exp\Bigl(-\int_0^\T \mathbf{S}(x^\dagger_t,\dot x^\dagger_t,D^\dagger_t,\dot D^\dagger_t)\,\dd t\Bigr)\,\rs(x_\T,D_\T) .
  \label{eq:pathprob-rev}
\end{equation}
Performing a change of integration variable $t\to\T-t$ in Eq.~\eqref{eq:pathprob-rev} yields
\begin{equation}
  P[\Path^\dagger] = \mathcal{N}\exp\Bigl(-\int_0^\T \mathbf{S}(x_t,-\dot x_t,D_t,-\dot D_t)\,\dd t\Bigr)\,\rs(x_\T,D_\T) .
  \label{eq:pathprob-rev-changed}
\end{equation}

\subsection{Detailed and integral fluctuation theorems}
\label{sec:total-DFT}

Substituting Eq.~\eqref{eq:OM} into the ratio of Eqs.~\eqref{eq:pathprob} and \eqref{eq:pathprob-rev-changed} yields the detailed fluctuation theorem (DFT) for the total entropy production at the trajectory level (see Appendix~\ref{app:total} for the explicit derivation),
\begin{equation}
  \frac{P[\Path]}{P[\Path^\dagger]} = \ee^{\Stot[\Path]},
  \label{eq:DFT-traj}
\end{equation}
with
\begin{equation}
  \Stot = \Smed + \Smf + \Ssys,
  \label{eq:Stot}
\end{equation}
where
\begin{subequations}
\label{eq:Stot-terms}
\begin{align}
  \Smed[\Path] &= \frac{1}{T}\int_0^\T \dot x_t\strat F(x_t,D_t)\,\dd t = -\frac{Q}{T}, \label{eq:Smed}\\
  \Smf[\Path]  &= \frac{2\tau_f}{\sigma^2}\int_0^\T \dot D_t\strat\bigl[O(x_t)-D_t\bigr]\,\dd t, \label{eq:Smf}\\
  \Ssys[\Path] &= \phi(x_\T,D_\T)-\phi(x_0,D_0). \label{eq:Ssys}
\end{align}
\end{subequations}
Here $Q=\int_0^\T \dot x_t\strat\partial_x[U+V]\,\dd t$ is the heat absorbed by the particle from the thermal bath~\cite{Sekimoto1997,Sekimoto2010}, and $\Ssys$ is the change in the stochastic Shannon entropy of the joint system. The term $\Smf$ has the form of the medium entropy production that would be associated with $D$ if the variable $D$ were to represent a physical degree of freedom in contact with a thermal bath. Here it quantifies the irreversibility associated with the dynamics of $D$ generated by the measurement and filtering process, and we label it accordingly (m\&f stands for measurement and filtering, rather than measurement and feedback). Note that it does not originate from the feedback, since it persists in the absence of feedback, i.e., $V(x,D)=0$.

Since each term in Eq.~\eqref{eq:Stot-terms} is odd under time reversal (Appendix~\ref{app:total-traj}), it follows that
\begin{equation}
    \Stot[\Path^\dagger]=-\Stot[\Path].
\end{equation}
Following the standard argument~\cite{Crooks2000,Seifert2005,Seifert2012}, which we reproduce in Appendix~\ref{app:general}, Eq.~\eqref{eq:DFT-traj} then implies the DFT and the integral fluctuation theorem (IFT) for the distribution of $\Stot$, and hence the second law,
\begin{equation}
  \frac{p(\Stot)}{p(-\Stot)} = \ee^{\Stot} \,\,,\,\,
  \avg{\ee^{-\Stot}} = 1 \,\,,\,\,
  \avg{\Stot}\ge 0 .
  \label{eq:DFT-tot}
\end{equation}

\section{Subsystem entropy production}
\label{sec:subsystem}

\subsection{Decomposition into subsystem contributions}
\label{sec:sub-decomp}

We now decompose the total entropy production into contributions associated with the dynamics of $x$ and $D$. Since $\Ssys$ depends only on the initial and final states, it can be written as the time integral of a total time derivative,
\begin{equation}
  \Ssys=-\int_0^\T \frac{\dd}{\dd t}\ln\rs(x_t,D_t)\,\dd t.
  \label{eq:time-integral}
\end{equation}
Applying the chain rule, which holds in the Stratonovich convention, to the integrand gives
\begin{equation}
  \frac{\dd}{\dd t}\ln\rs = \frac{1}{\rs}\Bigl(\dot x\strat\partial_x\rs + \dot D\strat\partial_D\rs + \partial_t\rs\Bigr).
  \label{eq:chain}
\end{equation}
The last term vanishes since $\rs$ is time-independent. Nonetheless, we keep it explicitly and use the Fokker-Planck equation~\eqref{eq:FP} to rewrite it as two current terms, which gives
\begin{equation}
  \frac{\dd}{\dd t}\ln\rs = \frac{1}{\rs}\Bigl(\dot x\strat\partial_x\rs - \partial_x\Jx\Bigr) + \frac{1}{\rs}\Bigl(\dot D\strat\partial_D\rs - \partial_D\JD\Bigr).
  \label{eq:chain-FP}
\end{equation}
Substituting Eqs.~\eqref{eq:time-integral} and \eqref{eq:chain-FP} into Eq.~\eqref{eq:Stot}, and grouping the terms associated with the dynamics of $x$ and of $D$, we obtain the decomposition
\begin{equation}
  \Stot = \Delta S_{x} + \Delta S_{D},
  \label{eq:split}
\end{equation}
where
\begin{subequations}
\label{eq:Spm}
\begin{align}
  \Delta S_{x}[\Gamma] &= \int_0^\T\Bigl[\dot x\strat\Bigl(\frac{F}{T} - \partial_x\ln\rs\Bigr) + \frac{\partial_x\Jx}{\rs}\Bigr]\dd t, \label{eq:Sxpm}\\
  \Delta S_{D}[\Gamma] &= \int_0^\T\Bigl[\dot D\strat\Bigl(\frac{2\tau_f(O-D)}{\sigma^2} - \partial_D\ln\rs\Bigr) + \frac{\partial_D\JD}{\rs}\Bigr]\dd t. \label{eq:SDpm}
\end{align}
\end{subequations}
We interpret $\Delta S_{x}$ and $\Delta S_{D}$ as the entropy production associated with the dynamics of the subsystems $x$ and $D$, respectively. Since $F/T=\mu_x/\Dx$ and $2\tau_f(O-D)/\sigma^2=\mu_D/\DD$, both expressions take the same form,
\begin{equation}
  \Delta S_{i}[\Gamma] = \int_0^\T\Bigl[\dot q_i\strat\Bigl(\frac{\mu_i}{\mathcal{D}_i}-\partial_i\ln\rs\Bigr)+\frac{\partial_iJ^s_i}{\rs}\Bigr]\dd t,
  \label{eq:S-general}
\end{equation}
with $i\in\{x,D\}$, $q_x = x$, and $q_D = D$. 

\subsection{Fluctuation theorems for subsystems}
\label{sec:sub-FT}

To derive fluctuation theorems for $\Delta S_{x}$ and $\Delta S_{D}$, we first recall the case of a prescribed protocol. Consider overdamped dynamics described by
\begin{equation}
  \dot x = f(x;\lambda)+\sqrt{2\mathcal{D}}\,\xi(t),
  \label{eq:driven}
\end{equation}
where the control parameter $\lambda$ follows a prescribed protocol $\lambda_t$. The corresponding adjoint dynamics, which relaxes to the same stationary distribution as the original dynamics but with the opposite stationary probability current, is obtained by applying the adjoint transformation $f\to\hat f = -f + 2\mathcal{D}\partial_x\ln\rs(x;\lambda)$~\cite{Chernyak2006,EspositoVdB2010PRL,VdBEsposito2010PRE}. Here and below, the hat denotes quantities associated with adjoint dynamics. Comparing the probability $P^F[\mathbf{X}]$ of a trajectory $\mathbf{X}=\{x_t\}$ under the forward process with the probability $\hat P^R[\mathbf{X}^\dagger]$ of the time-reversed trajectory under the adjoint dynamics with the reversed protocol leads to the fluctuation theorem for the non-autonomous entropy production~\cite{HatanoSasa2001,Chernyak2006,EspositoVdB2010PRL,VdBEsposito2010PRE} (Appendix~\ref{app:prescribed}),
\begin{equation}
  \frac{P^F[\mathbf{X}]}{\hat P^R[\mathbf{X}^\dagger]} = \ee^{\Delta S_{\mathrm{na}}[\mathbf{X}]},
  \label{eq:FT-na}
\end{equation}
with
\begin{equation}
  \Delta S_{\mathrm{na}}[\mathbf{X}] = -\int_0^\T \dot\lambda_t\,\partial_\lambda\ln\rs(x_t;\lambda_t)\,\dd t.
  \label{eq:na}
\end{equation}
This derivation, however, assumes that the protocol is prescribed.

In the present model, the feedback parameter $D$ is not prescribed. Instead, it is obtained from the measurement and filtering process, Eq.~\eqref{eq:EOMD}, and its realization depends on the trajectory of the particle $\{x_t\}$. Therefore, each realization $\{D_t\}$ can be regarded as a stochastic protocol. Unlike a prescribed protocol, a stochastic protocol is not under full control, and its time-reversed counterpart cannot be implemented manually. Instead, the time-reversed protocol corresponds to observing the time-reversed realization $\{D^\dagger_t\}$ under the same measurement and filtering process that generates $D$. The adjoint transformation should therefore be applied to the dynamics of $x$, but not to the measurement and filtering process, which yields the reference dynamics
\begin{subequations}
\label{eq:adj-x}
\begin{align}
  \dot x &= \hat\mu_x(x,D) + \sqrt{2\Dx}\,\xi_x, \\
  \dot D &= \mu_D(x,D) + \sqrt{2\DD}\,\xi_D ,
\end{align}
\end{subequations}
with $\hat\mu_x = -\mu_x + 2\Dx\partial_x\ln\rs$. This reference dynamics reverses the partial current of $x$,
\begin{equation}
  \hat J_x^s = \hat\mu_x\rs-\Dx\partial_x\rs = -\Jx ,
  \label{eq:Jx-hat}
\end{equation}
and leaves that of $D$ unchanged. Note that $\rs$ is not the stationary distribution of the reference dynamics unless $\partial_D\JD = 0$. Nonetheless, the reference dynamics defines a well-defined process, which is all that is required for deriving a fluctuation theorem (Appendix~\ref{app:sub}).

We let $\hat P_x[\Path^\dagger]$ denote the probability of observing the time-reversed trajectory $\Path^\dagger$ under the reference dynamics defined by Eq.~\eqref{eq:adj-x}, with the initial condition sampled from the stationary distribution of the original dynamics, $\rs(x_\T,D_\T)$. Evaluating the ratio of the corresponding path probabilities using the Onsager-Machlup functions (Appendix~\ref{app:sub}) gives the trajectory-level DFT
\begin{equation}
  \frac{P[\Path]}{\hat P_x[\Path^\dagger]} = \ee^{\Delta S_{D}[\Path]} .
  \label{eq:FT-D}
\end{equation}

Although the particle position $x$ is not physically a protocol, it formally plays the role of a stochastic protocol from the viewpoint of $D$, since the dynamics of $D$ depends on $x$ through $O(x)$. By symmetry, we therefore apply the adjoint transformation to the dynamics of $D$ only, while leaving the dynamics of $x$ unchanged. This yields the reference dynamics
\begin{subequations}
\label{eq:adj-D}
\begin{align}
  \dot x &= \mu_x(x,D) + \sqrt{2\Dx}\,\xi_x, \\
  \dot D &= \hat\mu_D(x,D) + \sqrt{2\DD}\,\xi_D ,
\end{align}
\end{subequations}
with $\hat\mu_D = -\mu_D + 2\DD\partial_D\ln\rs$. This reference dynamics reverses the partial current of $D$, $\hat J_D^s = \hat\mu_D\rs-\DD\partial_D\rs = -\JD$, and leaves that of $x$ unchanged; accordingly, $\rs$ is not its stationary distribution unless $\partial_x\Jx=0$. Denoting by $\hat P_D[\Path^\dagger]$ the probability of observing the time-reversed trajectory $\Path^\dagger$ under this reference dynamics, with the initial condition again sampled from $\rs(x_\T,D_\T)$, the same steps give
\begin{equation}
  \frac{P[\Path]}{\hat P_D[\Path^\dagger]} = \ee^{\Delta S_{x}[\Path]} .
  \label{eq:FT-x}
\end{equation}
Following the standard argument~\cite{Crooks2000,Seifert2005,Seifert2012} (see Appendices~\ref{app:general} and \ref{app:sub}), each of Eqs.~\eqref{eq:FT-D} and \eqref{eq:FT-x} implies the DFT for the distribution of $\Delta S_{i}$, the IFT, and a second-law-like inequality,
\begin{subequations}
\label{eq:IFT-sub}
\begin{align}
  \frac{p(\Delta S_{x})}{\hat p_D(-\Delta S_{x})} &= \ee^{\Delta S_{x}} \,\,,\,\,
  \avg{\ee^{-\Delta S_{x}}} = 1 \,\,,\,\,
  \avg{\Delta S_{x}}\ge 0,
  \label{eq:IFT-sub-x}\\
  \frac{p(\Delta S_{D})}{\hat p_x(-\Delta S_{D})} &= \ee^{\Delta S_{D}} \,\,,\,\,
  \avg{\ee^{-\Delta S_{D}}} = 1 \,\,,\,\,
  \avg{\Delta S_{D}}\ge 0.
  \label{eq:IFT-sub-D}
\end{align}
\end{subequations}
Here, $\hat p_D$ and $\hat p_x$ denote the distributions of $-\Delta S_{x}[\Path^\dagger]$ and $-\Delta S_{D}[\Path^\dagger]$, respectively, for trajectories $\Path$ generated by the corresponding reference dynamics. Note that
\begin{equation}
  -\Delta S_{i}[\Path^\dagger] = \int_0^\T\Bigl[\dot q_i\strat\Bigl(\frac{\mu_i}{\mathcal{D}_i}-\partial_i\ln\rs\Bigr)-\frac{\partial_iJ^s_i}{\rs}\Bigr]\dd t \neq \Delta S_{i}[\Path] ,
  \label{eq:S-general-minus}
\end{equation}
since the last term is even under time reversal. Hence, these distributions are in general not the distributions of $\Delta S_x$ and $\Delta S_D$ themselves.

Comparing the DFTs, Eqs.~\eqref{eq:FT-na} and \eqref{eq:FT-D}, we observe that they share the same structure. In both cases, the dynamics of $x$ is replaced by its adjoint in the reference process, and the time-reversed trajectory is considered. A key difference is that the time-reversed protocol $\{D^\dagger_t\}$ is produced by the measurement and filtering process, rather than implemented externally. Therefore, $\Delta S_{D}$ can be interpreted as the analog of the non-autonomous entropy production for the stochastic protocol.

In the decomposition $\Stot=\Delta S_{x}+\Delta S_{D}$, the complementary contribution $\Delta S_{x}$, associated with the dynamics of the subsystem $x$, then takes the place of the housekeeping entropy production $\Delta S_{\mathrm{hk}}=\Stot-\Delta S_{\mathrm{na}}$. Their fluctuation theorems, however, are derived from different reference dynamics. For $\Delta S_{\mathrm{hk}}$, the reference is the same trajectory under the adjoint dynamics of $x$, whereas for $\Delta S_{x}$, the reference is the time-reversed trajectory under the dynamics in which the measurement and filtering process is replaced by its adjoint. Although this construction is analogous to that for $\Delta S_{\mathrm{na}}$, with $x$ regarded as the protocol from the viewpoint of $D$, it has no physical counterpart in the case of a prescribed protocol, since $x$ is not physically a protocol. Moreover, $\Delta S_{x}$ does not behave like $\Delta S_{\mathrm{hk}}$. It does not quantify the entropy production arising from non-conservative forces, since it is in general nonzero even though the force acting on $x$ is conservative.

The fluctuation theorems in Eqs.~\eqref{eq:IFT-sub} coincide with the stationary case of the fluctuation theorems for the partial entropy production derived in Ref.~\cite{RosinbergHorowitz2016}. In Ref.~\cite{RosinbergHorowitz2016}, the reference dynamics was constructed specifically to facilitate the derivation of the desired fluctuation theorems. In our derivation, the same reference dynamics follows naturally from generalizing the derivation of the fluctuation theorem for $\Delta S_{\mathrm{na}}$, with the trajectory of the feedback parameter $D$ regarded as a stochastic protocol, and, by symmetry, with the trajectory of the particle position $x$ regarded as a stochastic protocol from the viewpoint of $D$.

\subsection{Relation to information flow}
\label{sec:sub-info}

The quantities $\Delta S_{x}$ and $\Delta S_{D}$ also admit an interpretation in terms of information thermodynamics~\cite{Allahverdyan2009,HorowitzEsposito2014,Parrondo2015}. Consider the stochastic mutual information between $x$ and $D$ in the stationary state,
\begin{equation}
  I^s_t = \ln\frac{\rs(x_t,D_t)}{\rs_x(x_t)\,\rs_D(D_t)},
  \label{eq:MI}
\end{equation}
where $\rs_x$ and $\rs_D$ are the marginal distributions. Applying the chain rule and the Fokker-Planck equation, Eq.~\eqref{eq:FP}, we decompose $\dot I^s_t=\dot I^s_x+\dot I^s_D$ into contributions from the dynamics of the subsystems $x$ and $D$,
\begin{subequations}
\label{eq:Idot}
\begin{align}
  \dot I^s_x &= \frac{\dot x\strat\partial_x\rs - \partial_x \Jx}{\rs} - \frac{\dd}{\dd t}\ln\rs_x, \label{eq:Idotx}\\
  \dot I^s_D &= \frac{\dot D\strat\partial_D\rs - \partial_D \JD}{\rs} - \frac{\dd}{\dd t}\ln\rs_D. \label{eq:IdotD}
\end{align}
\end{subequations}
Integrating $\dot I^s_x$ and $\dot I^s_D$ over $[0,\T]$ gives the stationary stochastic information flows $I^s_x$ and $I^s_D$~\cite{HorowitzEsposito2014,RosinbergHorowitz2016}. Comparing with Eq.~\eqref{eq:Spm}, we can rewrite $\Delta S_{x}$ and $\Delta S_{D}$ as
\begin{subequations}
\label{eq:S-info}
\begin{align}
  \Delta S_{x} &= \Smed + \Ssys^{x} - I^s_x, \label{eq:Sx-info}\\
  \Delta S_{D} &= \Smf + \Ssys^{D} - I^s_D, \label{eq:SD-info}
\end{align}
\end{subequations}
where $\Ssys^{x} = -\ln\rs_x(x_\T)+\ln\rs_x(x_0)$ and $\Ssys^{D} = -\ln\rs_D(D_\T)+\ln\rs_D(D_0)$ are the changes in the stochastic Shannon entropies of the subsystems.

This decomposition has a clear physical interpretation. If we consider only the particle, its apparent entropy production $\Smed+\Ssys^{x}$ does not satisfy a fluctuation theorem in general, due to the presence of measurement and feedback~\cite{SagawaUeda2012}. Including the information flow $I^s_x$ yields $\Delta S_{x}$, which does satisfy a fluctuation theorem. By symmetry, the same argument holds for the subsystem $D$. Combining Eqs.~\eqref{eq:IFT-sub} and \eqref{eq:S-info} gives the second-law-like inequalities
\begin{align}
  \avg{\Smed} + \avg{\Ssys^{x}} &\ge \avg{I^s_x}, \label{eq:2ndlaw-x}\\
  \avg{\Smf} + \avg{\Ssys^{D}} &\ge \avg{I^s_D}, \label{eq:2ndlaw-D}
\end{align}
in agreement with Refs.~\cite{Allahverdyan2009,HorowitzEsposito2014,HorowitzSandberg2014}. Since $\avg{\Ssys^{x}} = \avg{\Ssys^{D}} = 0$ in the stationary state, Eq.~\eqref{eq:2ndlaw-x} shows that the mean medium entropy production $\avg{\Smed}$ can be negative only if $\avg{I^s_x}$ is also negative. In other words, a net heat flow from the thermal bath into the particle is allowed only if information flows from the subsystem $x$ to the subsystem $D$, i.e., if $D$ acquires information about $x$ through the measurement.
\section{Solvable model: free particle with harmonic feedback}
\label{sec:harmonic}

We now consider the case of a free particle, $U(x)=0$, with measured observable $O(x)=x$, and a harmonic feedback potential of stiffness $k$ centered at the filtered signal $D$,
\begin{equation}
  V(x,D)=\frac{k}{2}(x-D)^2 .
  \label{eq:Vharm}
\end{equation}
The equations of motion become
\begin{subequations}
\label{eq:EOMharm}
\begin{align}
  \dot x &= -\frac{k}{\gamma}(x-D) + \sqrt{\frac{2T}{\gamma}}\,\xi_x,\\
  \dot D &= \frac{1}{\tau_f}(x-D) + \frac{\sigma}{\tau_f}\,\xi_D .
\end{align}
\end{subequations}

In the coordinates $w=\gamma x/k+\tau_f D$ and $y=x-D$, the equations of motion become
\begin{subequations}
\label{eq:wy}
\begin{align}
  \dot w &= \frac{\gamma}{k}\sqrt{\frac{2T}{\gamma}}\,\xi_x+\sigma\,\xi_D,\\
  \dot y &= -\Bigl(\frac{k}{\gamma}+\frac{1}{\tau_f}\Bigr)y + \sqrt{\frac{2T}{\gamma}}\,\xi_x - \frac{\sigma}{\tau_f}\,\xi_D .
\end{align}
\end{subequations}
We can see that $w$ undergoes free diffusion, while $y$ evolves autonomously as an Ornstein-Uhlenbeck process~\cite{UhlenbeckOrnstein1930}, although the noises acting on $w$ and $y$ are correlated. As a result, the joint process of $(x,D)$ does not admit a normalizable stationary distribution. Nonetheless, the corresponding Fokker-Planck equation admits the stationary solution
\begin{equation}
  \rs(x,D)\propto \ee^{-\beta^*V(x,D)} \,\,,\,\,
  \beta^* \equiv \frac{1}{T^*} = \frac{2\beta\tau_f(\gamma+k\tau_f)}{k(\beta\gamma\sigma^2+2\tau_f^2)} ,
  \label{eq:betastar}
\end{equation}
where $\beta=1/T$.

Although $\rs$ is not normalizable, it is sufficient for the analysis below. The relative coordinate $y$ relaxes to the normalizable stationary distribution
\begin{equation}
  \rs(y)=\sqrt{\frac{\beta^*k}{2\pi}}\,\ee^{-\beta^*ky^2/2} ,
  \label{eq:rho-y}
\end{equation}
with mean and variance
\begin{equation}
  \langle y\rangle = 0 \quad , \quad \langle y^2\rangle = \frac{1}{\beta^*k} .
  \label{eq:stat-y}
\end{equation}
Since the heat and the work [Eqs.~\eqref{eq:QW} below] depend only on $y$ and the noise terms, their stationary averages can be evaluated with respect to $\rs(y)$. Moreover, the subsystem entropy productions, Eqs.~\eqref{eq:Spm}, depend on $\rs$ only through $\partial_i\ln\rs$ and $\partial_iJ^s_i/\rs$, which are independent of the normalization of $\rs$. For $\rs\propto\ee^{-\beta^*V}$, these terms take the explicit forms
\begin{subequations}
\label{eq:harm-derivs}
\begin{align}
  \partial_x\ln\rs&=-\partial_D\ln\rs=-\beta^*ky, \label{eq:harm-derivs-log}\\
  \frac{\partial_x\Jx}{\rs}&=-\frac{\partial_D\JD}{\rs}=\frac{k}{\gamma\beta}(\beta^*-\beta)\bigl(1-\beta^*ky^2\bigr). \label{eq:harm-derivs-J}
\end{align}
\end{subequations}
Hence, $\Delta S_{x}$ and $\Delta S_{D}$ are well defined and, after substituting the equations of motion for $\dot x$ and $\dot D$, can be expressed in terms of $y$ and the noise.

\subsection{Energetics and feedback cooling}
\label{sec:harm-energetics}

We first consider the energetics of the model. The internal energy of the particle is given by $V(x,D)$, and its rate of change obeys the first law of thermodynamics, $\dot V=\dot Q+\dot W$, with
\begin{subequations}
\label{eq:QW-def}
\begin{align}
  \dot Q &= \dot x\strat\partial_xV = k\,\dot x\strat y ,\\
  \dot W &= \dot D\strat\partial_DV = -k\,\dot D\strat y .
\end{align}
\end{subequations}
Here, $\dot Q$ is the rate of heat absorbed from the thermal bath, and $\dot W$ is the rate of work performed by the feedback~\cite{Sekimoto1997}. In the It\^o convention, their explicit expressions are given by (Appendix~\ref{app:energetics})
\begin{subequations}
\label{eq:QW}
\begin{align}
  \dot Q &= -\frac{k^2}{\gamma}y^2 + \frac{kT}{\gamma} + ky\sqrt{\frac{2T}{\gamma}}\,\xi_x ,\\
  \dot W &= -\frac{k}{\tau_f}y^2 + \frac{k\sigma^2}{2\tau_f^2} - \frac{k\sigma}{\tau_f}\,y\,\xi_D .
\end{align}
\end{subequations}
Using Eq.~\eqref{eq:stat-y} and the fact that $\langle y\,\xi_x\rangle=\langle y\,\xi_D\rangle=0$ in the It\^o convention, we obtain the stationary means
\begin{equation}
  \langle\dot Q\rangle^s = -\langle\dot W\rangle^s = -\frac{k(k\sigma^2-2T\tau_f)}{2\tau_f(\gamma+k\tau_f)},
  \label{eq:Qmean}
\end{equation}
which is consistent with the first law in the stationary state, $\langle\dot Q\rangle^s+\langle\dot W\rangle^s=0$. For performing feedback cooling, i.e., $\langle\dot Q\rangle^s>0$, it therefore requires that the parameters to satisfy the criteria
\begin{equation}
  T > \frac{k\sigma^2}{2\tau_f}.
  \label{eq:cooling}
\end{equation}
In this regime, heat flows continuously from the thermal bath into the particle and is then extracted by the feedback. For a fixed $T$, the cooling regime can be reached by lowering the feedback stiffness $k$, reducing the measurement noise amplitude $\sigma$, or increasing the filter time constant $\tau_f$.

\subsection{Entropy production and information flow}
\label{sec:harm-entropy}

We now move on to consider the entropy production of the model. Substituting Eqs.~\eqref{eq:harm-derivs} into Eqs.~\eqref{eq:Spm} and using the definitions of $\dot Q$ and $\dot W$, Eqs.~\eqref{eq:QW-def}, the subsystem entropy production rates take the compact form
\begin{subequations}
\label{eq:rates-harm}
\begin{align}
  \Delta\dot S_{x} &= (\beta^*-\beta)\Bigl[\dot Q + \frac{k}{\gamma\beta}\bigl(1-\beta^* k y^2\bigr)\Bigr], \label{eq:Sxdot}\\
  \Delta\dot S_{D} &= \Bigl(\beta^*-\frac{2\tau_f}{k\sigma^2}\Bigr)\Bigl[\dot W + \frac{k\sigma^2}{2\tau_f^2}\bigl(1-\beta^* k y^2\bigr)\Bigr]. \label{eq:SDdot}
\end{align}
\end{subequations}
The terms proportional to $1-\beta^*ky^2$ have zero stationary mean, since $k\langle y^2\rangle=1/\beta^*$ by Eq.~\eqref{eq:stat-y}. Using Eqs.~\eqref{eq:betastar} and \eqref{eq:Qmean}, the stationary means are given by (Appendix~\ref{app:harm-entropy})
\begin{subequations}
\label{eq:rates-mean}
\begin{align}
  \langle\Delta\dot S_{x}\rangle^s &= (\beta^*-\beta)\langle\dot Q\rangle^s \nonumber\\
   &= \frac{\gamma(\beta k\sigma^2-2\tau_f)^2}{2\tau_f(\beta\gamma\sigma^2+2\tau_f^2)(\gamma+k\tau_f)} \ge 0, \label{eq:Sxmean}\\
  \langle\Delta\dot S_{D}\rangle^s &= \Bigl(\frac{2\tau_f}{k\sigma^2}-\beta^*\Bigr)\langle\dot Q\rangle^s \nonumber\\
   &= \frac{\tau_f(\beta k\sigma^2-2\tau_f)^2}{\beta\sigma^2(\beta\gamma\sigma^2+2\tau_f^2)(\gamma+k\tau_f)} \ge 0, \label{eq:SDmean}
\end{align}
\end{subequations}
which verifies the second-law-like inequalities explicitly. Their sum gives the total entropy production rate,
\begin{equation}
  \langle\Delta\dot S_{\mathrm{tot}}\rangle^s = \Bigl(\frac{2\tau_f}{k\sigma^2}-\beta\Bigr)\langle\dot Q\rangle^s = \frac{(\beta k\sigma^2-2\tau_f)^2}{2\beta\tau_f\sigma^2(\gamma+k\tau_f)}\ge0 .
  \label{eq:Stotmean}
\end{equation}
All three entropy production rates vanish if and only if $T=k\sigma^2/(2\tau_f)$. In that case, $\beta^*=\beta$, and the stationary state is an equilibrium state.

Finally, the stationary information flows follow from Eqs.~\eqref{eq:S-info}. Since the mean rates of change of the subsystem Shannon entropies vanish in the stationary state, we have
\begin{subequations}
\label{eq:Iflow-S}
\begin{align}
  \langle\dot I^s_x\rangle^s &= \langle\Delta\dot S_{\mathrm{med}}\rangle^s-\langle\Delta\dot S_{x}\rangle^s, \\
  \langle\dot I^s_D\rangle^s &= \langle\Delta\dot S_{\mathrm{m\&f}}\rangle^s-\langle\Delta\dot S_{D}\rangle^s.
\end{align}
\end{subequations}
For the present model, Eqs.~\eqref{eq:Smed} and \eqref{eq:Smf}, together with the definitions of $\dot Q$ and $\dot W$, give
\begin{equation}
  \Delta\dot S_{\mathrm{med}} = -\beta\dot Q \quad , \quad
  \Delta\dot S_{\mathrm{m\&f}} = -\frac{2\tau_f}{k\sigma^2}\dot W .
  \label{eq:Smed-Smf-harm}
\end{equation}
Substituting Eqs.~\eqref{eq:Qmean}, \eqref{eq:rates-mean}, and \eqref{eq:Smed-Smf-harm} into Eqs.~\eqref{eq:Iflow-S}, we obtain
\begin{equation}
  \langle\dot I^s_x\rangle^s = -\langle\dot I^s_D\rangle^s = -\beta^*\langle\dot Q\rangle^s.
  \label{eq:Iflow-harm}
\end{equation}

\subsection{Mapping onto a two-temperature system}
\label{sec:harm-mapping}

These results admit an intuitive thermodynamic interpretation through a mapping onto an autonomous system. Consider an autonomous bipartite system as illustrated by Fig.~\ref{fig:schematic}(b), in which both $x$ and $D$ represent physical degrees of freedom. The variable $x$ is coupled to a thermal bath with friction coefficient $\gamma$ and temperature $T$, while the variable $D$ is coupled to a second thermal bath with friction coefficient $\gamma_D$ and temperature $T_D$, and the two interact through the potential $V=k(x-D)^2/2$. The equations of motion are
\begin{subequations}
\label{eq:autonomous}
\begin{align}
  \dot x &= -\frac{1}{\gamma}\partial_xV + \sqrt{\frac{2T}{\gamma}}\,\xi_x ,\\
  \dot D &= -\frac{1}{\gamma_D}\partial_DV + \sqrt{\frac{2T_D}{\gamma_D}}\,\xi_D .
\end{align}
\end{subequations}
By choosing
\begin{equation}
  \gamma_D = k\tau_f \,\,,\,\, T_D = \frac{k\sigma^2}{2\tau_f},
  \label{eq:TD}
\end{equation}
Eqs.~\eqref{eq:autonomous} coincide with Eqs.~\eqref{eq:EOMharm}. The harmonic feedback model is therefore mathematically equivalent to two coupled overdamped degrees of freedom in contact with thermal baths at different temperatures, similar to the model studied in Ref.~\cite{LoosKlapp2020}. In this picture, the filtered signal $D$ plays the role of a physical degree of freedom in contact with a thermal bath whose temperature and friction coefficient are determined by the feedback stiffness, the filter time constant, and the measurement noise through Eq.~\eqref{eq:TD}. The effective temperature $T^*$ in Eq.~\eqref{eq:betastar} then takes the form of an inverse-friction-weighted average of the two bath temperatures,
\begin{equation}
  T^* = \frac{\gamma^{-1} T + \gamma_D^{-1} T_D}{\gamma^{-1}+\gamma_D^{-1}} = \frac{\gamma_D T + \gamma T_D}{\gamma+\gamma_D}.
  \label{eq:Tstar}
\end{equation}

This equivalence gives every quantity derived above a direct thermodynamic meaning. (i) Since $D$ is now a physical degree of freedom, $\dot W=\dot D\strat\partial_DV$ is the rate of heat absorbed by $D$ from the second bath. By Eqs.~\eqref{eq:Smed-Smf-harm} and \eqref{eq:TD}, $\Delta\dot S_{\mathrm{m\&f}}=-\dot W/T_D$ is then precisely the medium entropy production rate of that bath. (ii) The feedback-cooling condition, Eq.~\eqref{eq:cooling}, becomes $T>T_D$. In this regime, heat flows down the temperature gradient from the hotter physical bath toward the colder effective bath associated with measurement, filtering, and feedback. (iii) In the stationary state, $x$ absorbs heat at a rate $\langle\dot Q\rangle^s$ from its bath and passes this heat to $D$, which in turn releases it into the second bath. (iv) The total entropy production rate, $\langle\Delta\dot S_{\mathrm{tot}}\rangle^s=(\beta_D-\beta)\langle\dot Q\rangle^s$ with $\beta_D = 1/T_D$, arises from the steady heat conduction between the two baths. The subsystem contributions, $\langle\Delta\dot S_{x} \rangle^s=(\beta^*-\beta)\langle\dot Q\rangle^s$ and $\langle\Delta\dot S_{D}\rangle^s = (\beta_D-\beta^*)\langle\dot Q\rangle^s$, take the form as if the joint system of $x$ and $D$ were held at the effective inverse temperature $\beta^*$, consistent with the Boltzmann form stationary distribution $\rs\propto\ee^{-\beta^*V}$. (v) Correspondingly, the information flow $\langle\dot I^s_D\rangle^s=-\langle\dot I^s_x\rangle^s=\beta^*\langle\dot Q\rangle^s$ can be interpreted as the entropy rate associated with the heat conduction between $x$ and $D$ at temperature $T^*$.

We emphasize that the mapping holds at the level of mathematical equivalence, Eqs.~\eqref{eq:EOMharm} and \eqref{eq:autonomous}. The effective bath with temperature $T_D$ and friction coefficient $\gamma_D$ is a property of the measurement, filtering, and feedback process, and $\Smf$ does not account for the thermodynamic cost of operating the measurement, filtering, and feedback devices. Related effective descriptions of feedback cooling with noisy, filtered measurements have been discussed in Refs.~\cite{MunakataRosinberg2013,HorowitzSandberg2014,RosinbergHorowitz2016}.

\subsection{Numerical verification}
\label{sec:harmonic-numerics}

We performed numerical simulations to verify the analytical results above, as well as the fluctuation theorems of Secs.~\ref{sec:total} and \ref{sec:subsystem}. Trajectories were generated from the equations of motion, Eqs.~\eqref{eq:EOMharm}, using the Euler--Maruyama method~\cite{Maruyama1955}. The relative coordinate $y$ was assumed to be in its stationary state at $t=0$; therefore, the initial value of $y$ was sampled from $\rs(y)$. The initial value of $w$, which we set to zero, is irrelevant, as all quantities computed here are independent of $w$. Along each trajectory, $Q$, $W$, $\Delta S_{x}$, and $\Delta S_{D}$ were computed by integrating the corresponding rates, Eqs.~\eqref{eq:QW} and \eqref{eq:rates-harm}, over time. The reported statistical errors are standard errors obtained by a Poisson bootstrap with 2000 replicates.

\begin{figure}[!hbtp]
   \centering
  \begin{tabular}{@{}l@{\hspace{0.02\linewidth}}l@{}}
    \small(a) & \small(b) \\[-1pt]
    \includegraphics[width=0.49\linewidth]{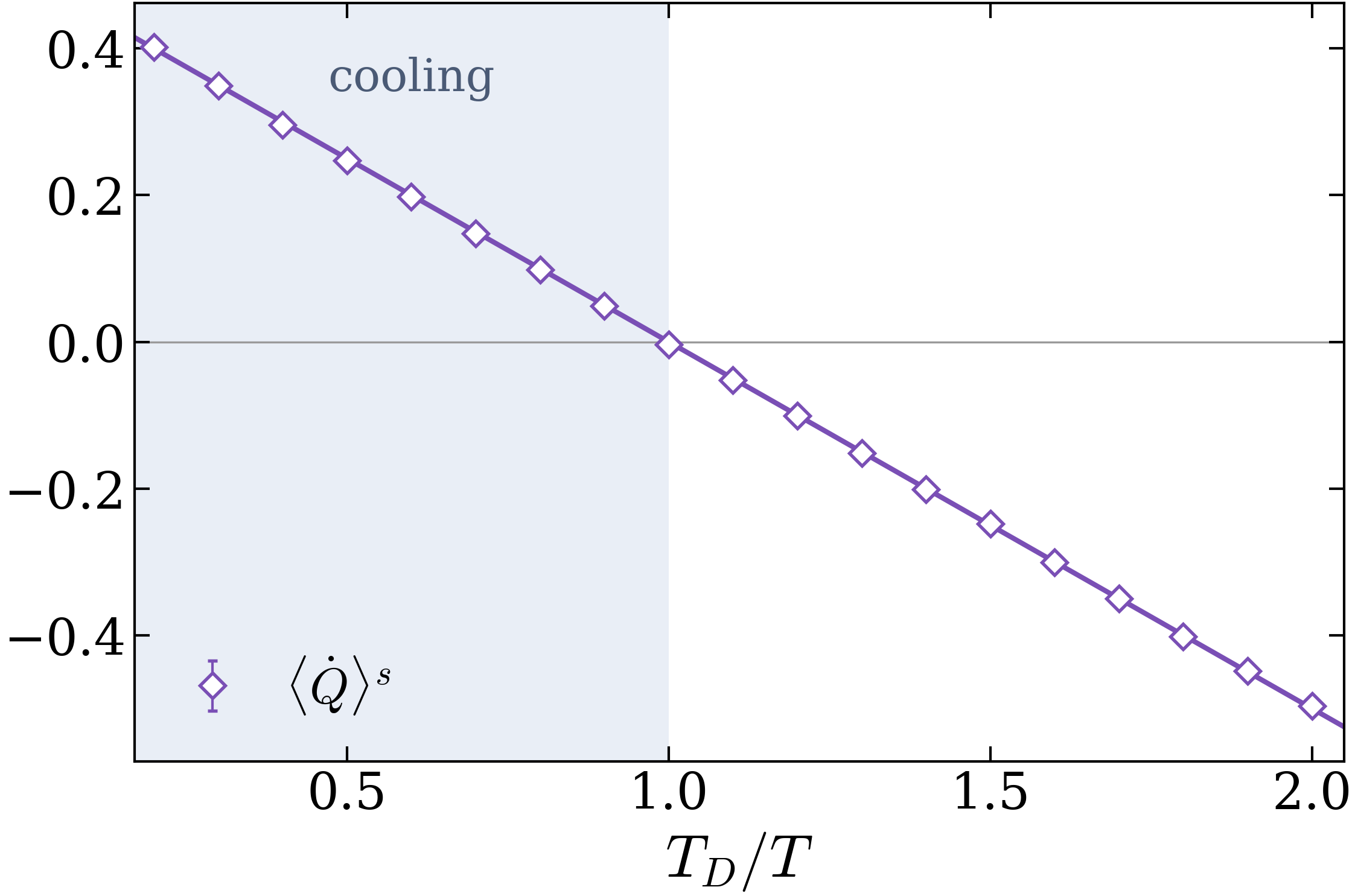} &
    \includegraphics[width=0.49\linewidth]{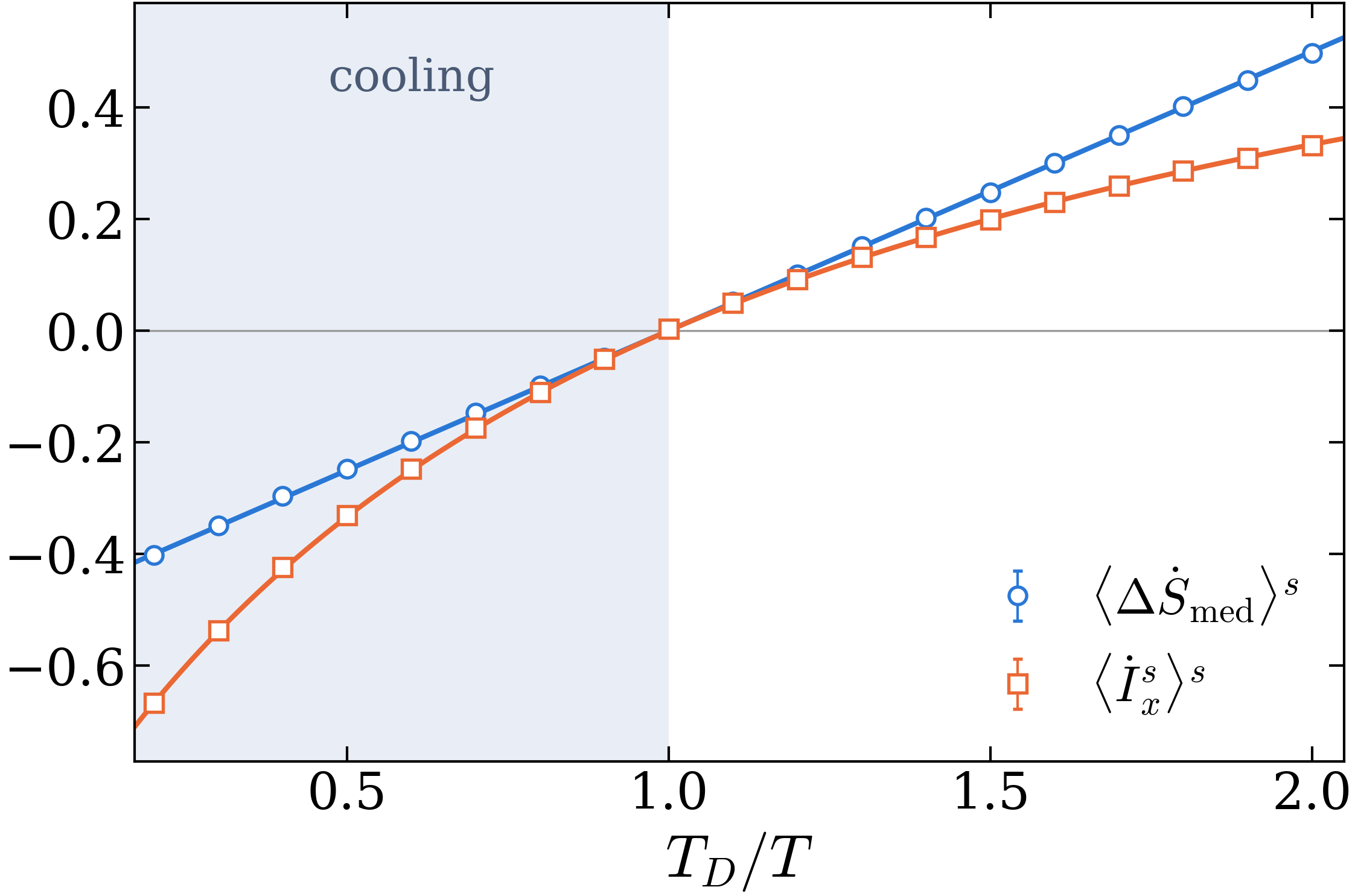} \\[2pt]
    \small(c) & \small(d) \\[2pt]
    \includegraphics[width=0.49\linewidth]{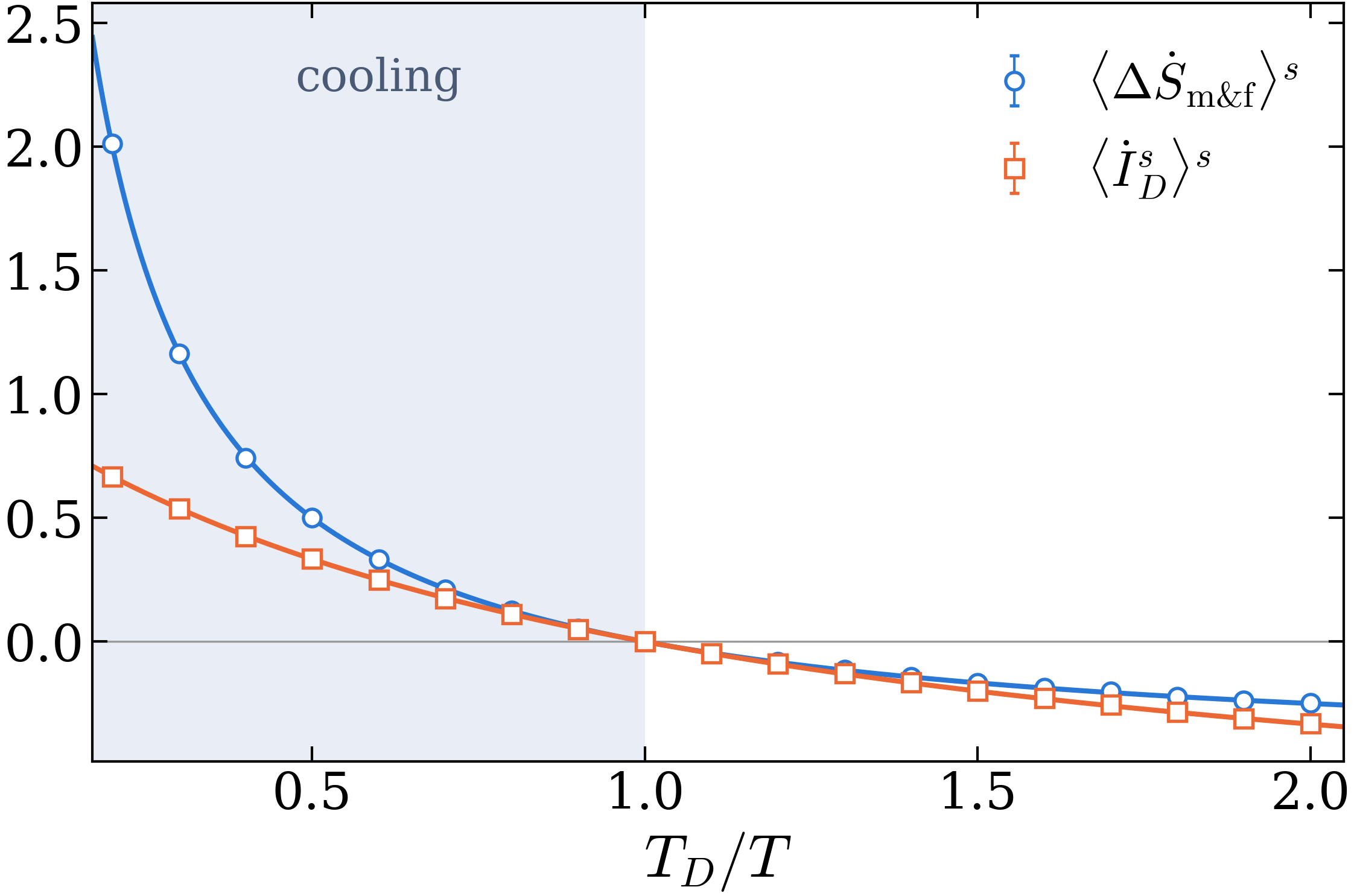} &
    \includegraphics[width=0.49\linewidth]{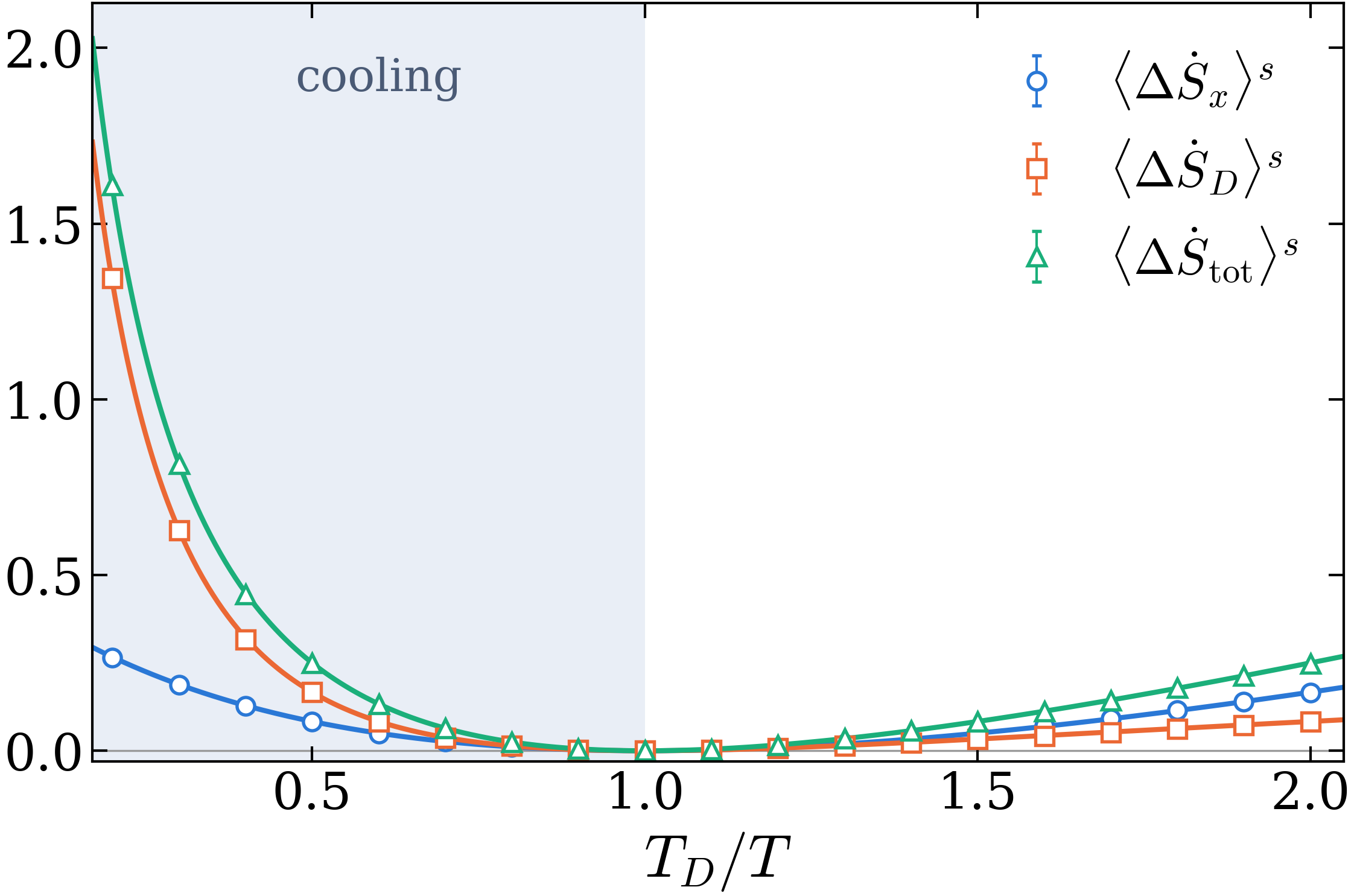}
  \end{tabular}
    \caption{Stationary mean rates of the harmonic feedback model as functions of $T_D/T$.
  (a) Heat absorbed by the particle from its bath.
  (b) Subsystem $x$: medium entropy production and information flow.
  (c) Subsystem $D$: measurement-and-filtering entropy term and information flow.
  (d) Subsystem and total entropy production rates.
  Solid lines: theoretical predictions, Eqs.~\eqref{eq:Qmean}, \eqref{eq:rates-mean}, \eqref{eq:Stotmean}, \eqref{eq:Smed-Smf-harm}, and \eqref{eq:Iflow-harm}; symbols: simulations. The shaded region indicates the cooling regime $T_D<T$. Parameters: $k=\gamma=T=\tau_f=1$, with $\sigma$ chosen such that $T_D/T$ ranges from $0.2$ to $2.0$. Each point is obtained from $N=10^5$ trajectories of duration $\T=2$, with time step $\mathrm{d}t=10^{-3}$. Error bars are smaller than the symbols.}
  \label{fig:rates}
\end{figure}

Figure~\ref{fig:rates} shows the stationary mean rates as functions of $T_D/T$, where $T_D$ is varied through the measurement noise $\sigma$ alone. The simulated rates agree with the theory over the whole range, and the error bars are much smaller than the symbols. Panel (a) confirms that the particle absorbs heat from its bath, $\langle\dot Q\rangle^s>0$, only for $T_D<T$, in agreement with the feedback-cooling condition derived in Sec.~\ref{sec:harm-energetics}.

Panels (b) and (c) show the second-law-like inequalities for the two subsystems, Eqs.~\eqref{eq:2ndlaw-x} and \eqref{eq:2ndlaw-D}, in which the mean rates of change of the system entropy vanish in the stationary state, $\langle\Delta\dot S^x_{\mathrm{sys}}\rangle^s=\langle\Delta\dot S^D_{\mathrm{sys}}\rangle^s=0$. In both panels, the entropy terms lie above the corresponding information flows, $\langle\Delta\dot S_{\mathrm{med}}\rangle^s\ge\langle\dot I^s_x\rangle^s$ and $\langle\Delta\dot S_{\mathrm{m\&f}}\rangle^s\ge\langle\dot I^s_D\rangle^s$, over the tested range, including the cooling regime, where $\langle\Delta\dot S_{\mathrm{med}}\rangle^s<0$. The gap between the two curves in each panel is the corresponding subsystem entropy production, which is shown in panel (d) together with the total entropy production rate. All entropy production rates are nonnegative and vanish at $T_D=T$, where the joint system is in equilibrium.

To test the fluctuation theorems, we generated three independent ensembles of $N=10^7$ trajectories each, with the forward, $x$-adjoint, and $D$-adjoint dynamics, described by Eqs.~\eqref{eq:EOMharm}, \eqref{eq:adj-x}, and \eqref{eq:adj-D}, respectively. We used $k=2$, $\gamma=1$, $T=0.5$, $\sigma=1.5$, and $\tau_f=2$, i.e., $T_D=1.125$, $\gamma_D=4$, and $T^*=0.625$. Each trajectory has duration $\T=2$ with an integration time step $\mathrm{d}t = 10^{-4}$. With this parameter choice, the typical entropy production remains of order unity, such that trajectories with negative entropy production, which are important for verifying fluctuation theorems, can be well sampled with a reasonable number of realizations. Table~\ref{tab:means} lists the sample means over the time interval $[0,\T]$ obtained from the forward ensemble with these parameters; they all agree with the theoretical predictions to within $2$ standard errors, and $\langle Q\rangle^s+\langle W\rangle^s=-0.0003\pm0.0005$, consistent with the energy balance predicted by the first law in the stationary state. The IFTs, Eqs.~\eqref{eq:DFT-tot} and \eqref{eq:IFT-sub}, are also satisfied to within statistical errors.

\begin{table}[!hbtp]
  \caption{Sample means over $[0,\T]$ for the forward ensemble, compared with the theoretical predictions, given by $\T$ times the stationary mean rates in Eqs.~\eqref{eq:Qmean}--\eqref{eq:Stotmean}, and with the IFTs, Eqs.~\eqref{eq:DFT-tot} and \eqref{eq:IFT-sub}. Errors are one standard error.}
  \label{tab:means}
  \begin{ruledtabular}
  \begin{tabular}{lcc}
    Quantity & Simulation & Theory \\
    \hline
    $\langle Q\rangle^s$ & $-0.5002\pm0.0003$ & $-0.5$ \\
    $\langle W\rangle^s$ & $0.4999\pm0.0003$ & $0.5$ \\
    $\langle\Delta S_{x}\rangle^s$ & $0.2003\pm0.0002$ & $0.2$ \\
    $\langle\Delta S_{D}\rangle^s$ & $0.3552\pm0.0002$ & $0.3556$ \\
    $\langle\Delta S_{\mathrm{tot}}\rangle^s$ & $0.5555\pm0.0003$ & $0.5556$ \\
    \hline
    $\langle\ee^{-\Delta S_{x}}\rangle$ & $0.9997\pm0.0002$ & $1$ \\
    $\langle\ee^{-\Delta S_{D}}\rangle$ & $0.9992\pm0.0008$ & $1$ \\
    $\langle\ee^{-\Delta S_{\mathrm{tot}}}\rangle$ & $1.0003\pm0.0005$ & $1$ \\
  \end{tabular}
  \end{ruledtabular}
\end{table}

\begin{figure}[!hbtp]
  \centering
  \includegraphics[width=\linewidth]{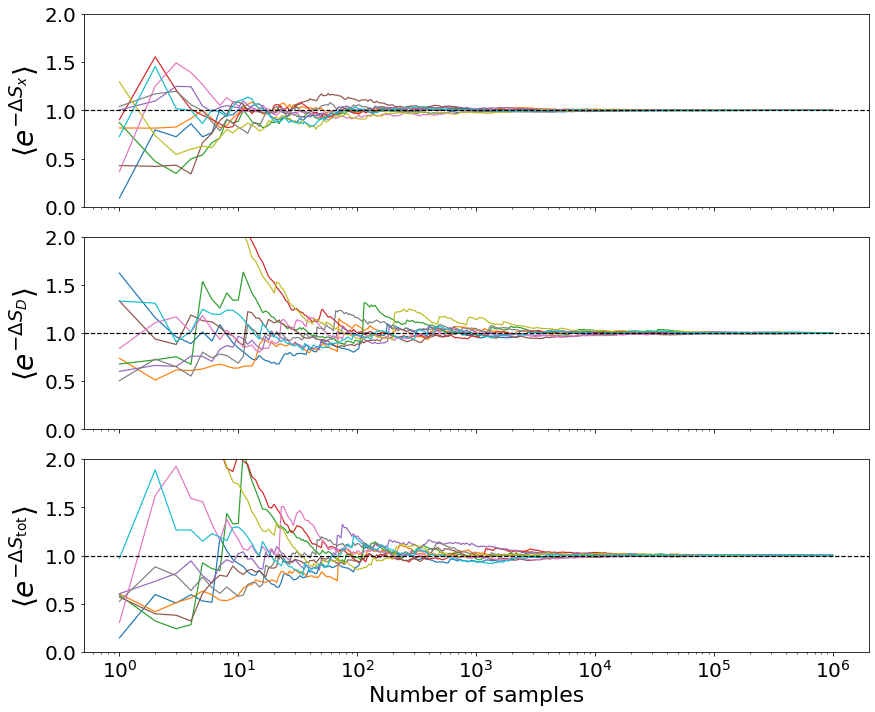}
  \caption{Convergence of the integral fluctuation theorems. Running averages of $\ee^{-\Delta S}$ versus the number of trajectories for (top) $\Delta S_{x}$, (middle) $\Delta S_{D}$, and (bottom) $\Delta S_{\mathrm{tot}}$, computed from ten independent batches of $10^6$ forward trajectories (one color per batch).}
  \label{fig:ift}
\end{figure}

Since the exponential average is dominated by rare trajectories with negative entropy production, we also checked the convergence of the IFTs. Figure~\ref{fig:ift} shows running averages of $\ee^{-\Delta S}$ for ten independent batches of $10^6$ trajectories. For $\Delta S_{x}$, the running averages of all batches collapse onto each other and onto unity on the scale of Fig.~\ref{fig:ift} after about $10^3$ trajectories, whereas for $\Delta S_{D}$ and $\Delta S_{\mathrm{tot}}$ this takes about $10^4$ trajectories. The results in Table~\ref{tab:means}, obtained from $10^7$ trajectories, are therefore well converged.

\begin{figure*}[!htbp]
  \centering
  \begin{tabular}{@{}l@{\hspace{0.01\linewidth}}l@{\hspace{0.01\linewidth}}l@{}}
    \small(a) & \small(b) & \small(c) \\[-1pt]
    \includegraphics[width=0.32\linewidth]{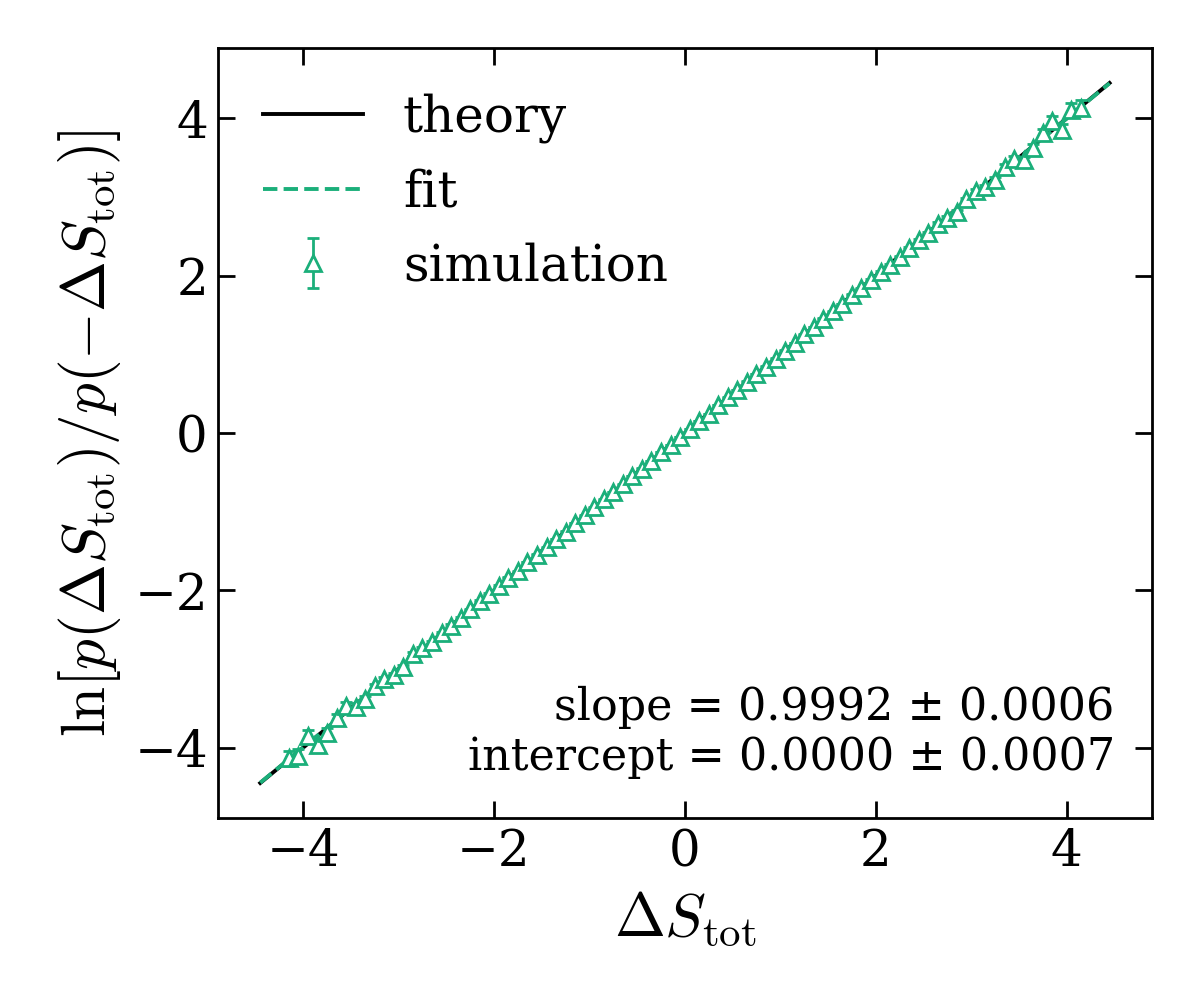} &
    \includegraphics[width=0.32\linewidth]{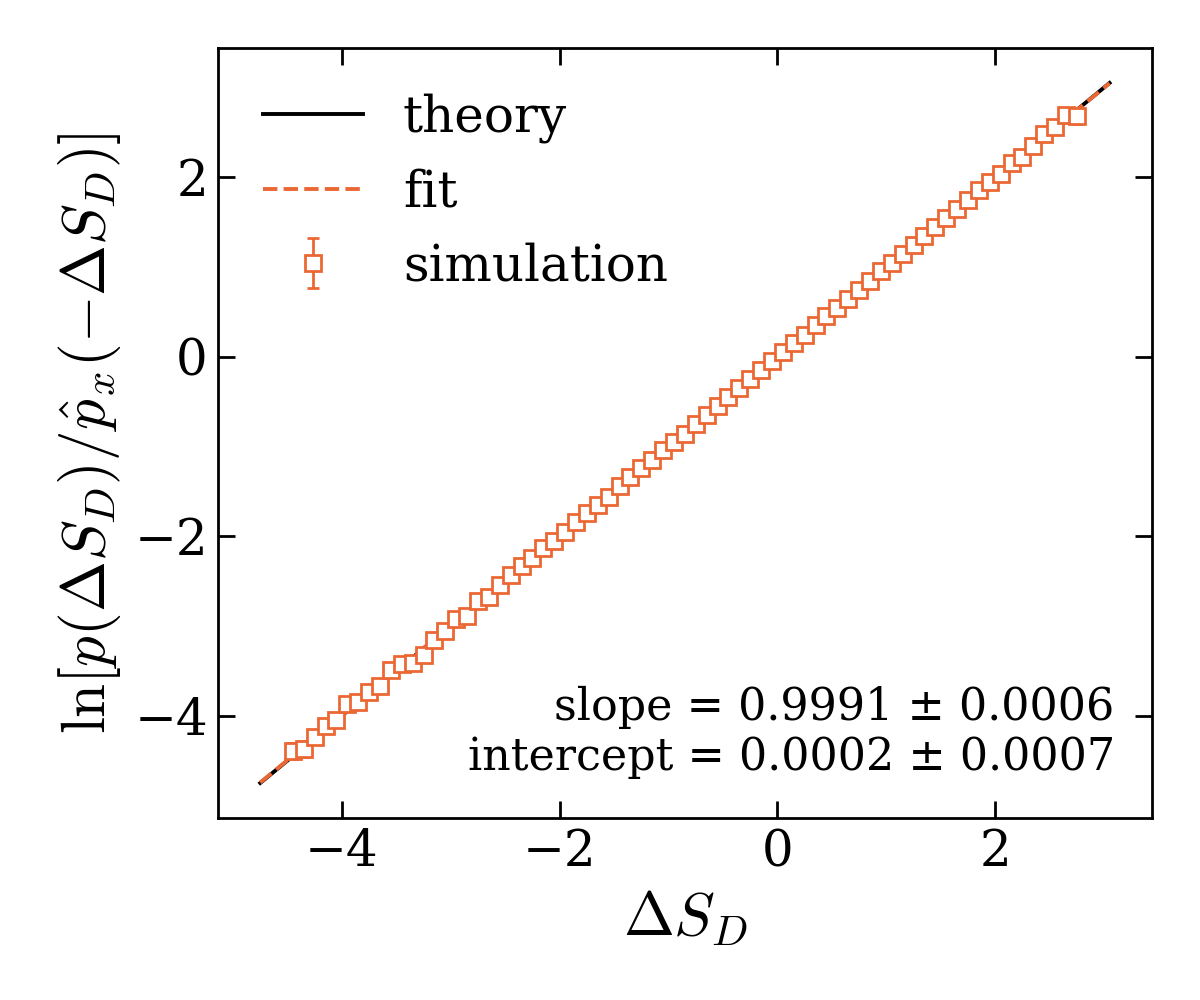} &
    \includegraphics[width=0.32\linewidth]{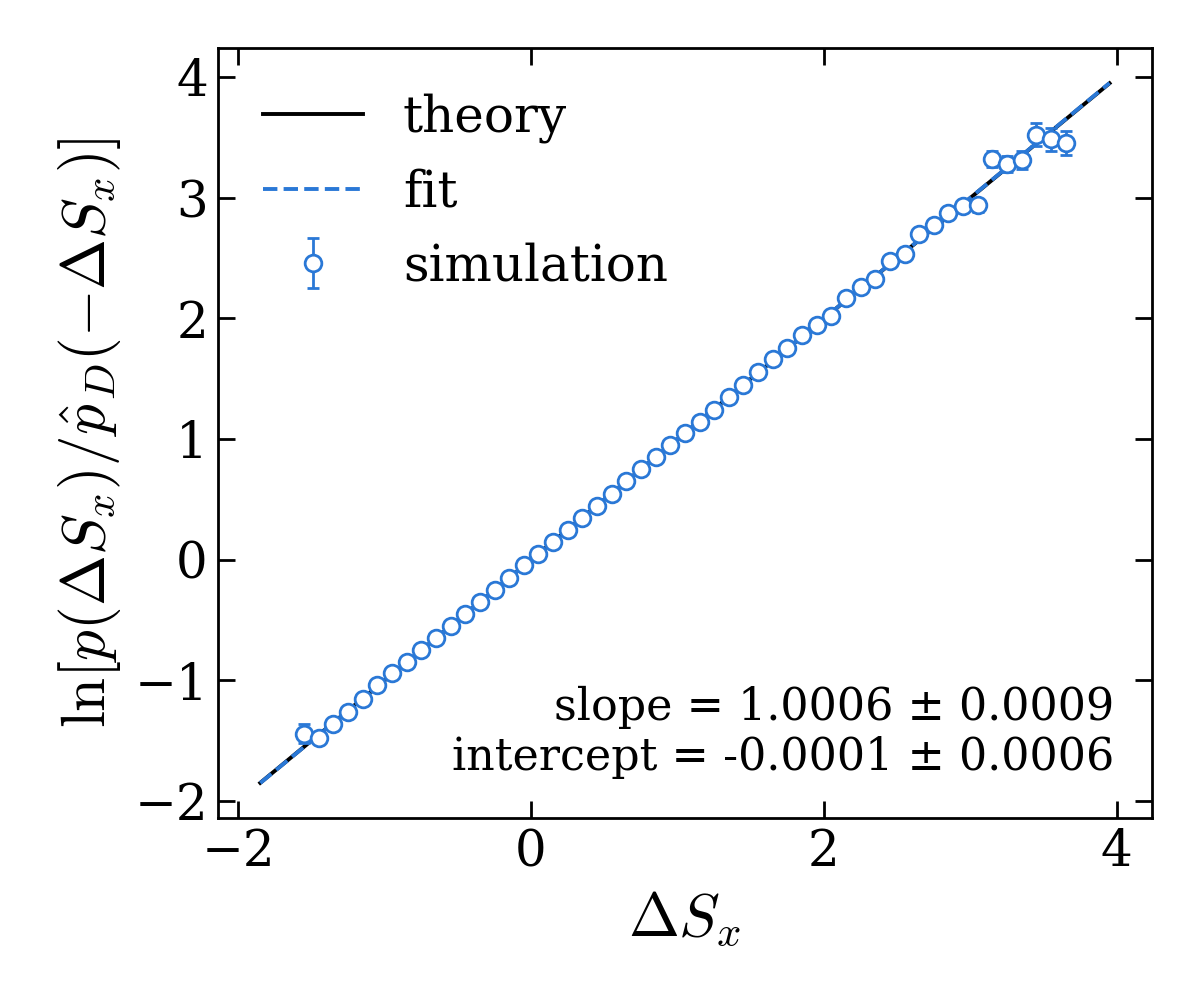}
  \end{tabular}
  \caption{Detailed fluctuation theorems for (a) $\Delta S_{\mathrm{tot}}$, (b) $\Delta S_{D}$, and (c) $\Delta S_{x}$. Symbols: simulated log-ratios $\ln[p(s)/p_{\mathrm{ref}}(-s)]$ from histograms with bin width $0.1$, shown for bins with at least $100$ counts in both histograms; error bars from bootstrap. Solid lines: theory, Eq.~\eqref{eq:DFT-log}. Dashed lines: fits. Parameters: $k=2$, $\gamma=1$, $T=0.5$, $\sigma=1.5$, $\tau_f=2$, $\T=2$, $\mathrm{d}t=10^{-4}$, and $N=10^7$ trajectories per ensemble.}
  \label{fig:dft}
\end{figure*}
 
To verify the DFTs, Eqs.~\eqref{eq:DFT-tot}, \eqref{eq:IFT-sub-x}, and \eqref{eq:IFT-sub-D}, we take the logarithm of both sides. The DFTs then state that the log-ratio of the forward and reference distributions is linear in the entropy production,
\begin{equation}
  \ln\frac{p(s)}{p_{\mathrm{ref}}(-s)}=s,
  \label{eq:DFT-log}
\end{equation}
where $p_{\mathrm{ref}}=\hat p_x$ for $s=\Delta S_{D}$, $p_{\mathrm{ref}}=\hat p_D$ for $s=\Delta S_{x}$, and $p_{\mathrm{ref}}=p$ for $s=\Delta S_{\mathrm{tot}}$. The reference distributions $\hat p_x$ and $\hat p_D$ were obtained from the $x$-adjoint and $D$-adjoint ensembles, respectively. We therefore fitted the simulated log-ratios to the linear function $as+c$, for which the theory predicts $a=1$ and $c=0$, using a binomial maximum-likelihood fit to the histogram counts. We obtain
\begin{equation}
\begin{aligned}
  \Delta S_{\mathrm{tot}}:&\quad a=0.9992\pm0.0006,\quad c=0.0000\pm0.0007,\\
  \Delta S_{D}:&\quad a=0.9991\pm0.0006,\quad c=0.0002\pm0.0007,\\
  \Delta S_{x}:&\quad a=1.0006\pm0.0009,\quad c=-0.0001\pm0.0006,
\end{aligned}
\label{eq:DFT-fit}
\end{equation}
all within $2$ standard errors of the theoretical values (Fig.~\ref{fig:dft}).

\section{Conclusion}
\label{sec:conclusion}

In this paper, we investigated how the decomposition of the entropy production into the housekeeping and non-autonomous contributions and the corresponding fluctuation theorems change when the protocol is not prescribed but determined by the measurement and feedback process. We considered an overdamped Brownian particle under continuous feedback control with a low-pass-filtered measurement signal $D$, and regarded each realization of $D$ as a stochastic protocol. We decomposed the total entropy production into two contributions, $\Delta S_x$ and $\Delta S_D$, associated with the dynamics of the particle and of the filtered signal, respectively. By extending the derivation for a prescribed protocol to a stochastic protocol, we derived fluctuation theorems for both contributions, which coincide with the stationary case of those in Ref.~\cite{RosinbergHorowitz2016}. We found that $\Delta S_D$ can be interpreted as the analog of the non-autonomous entropy production for a stochastic protocol. In contrast, $\Delta S_x$ is not the analog of the housekeeping entropy production, as it does not vanish in the absence of non-conservative forces. Both contributions can also be expressed in terms of the information flow, which quantifies the information exchanged between the two subsystems. Through the measurement, the filtered signal obtains information about the particle, which is then used to implement feedback.

We then studied a solvable model of a free particle with harmonic feedback, for which we obtained explicit expressions for all relevant quantities. We showed that this model can be mapped exactly onto an autonomous system of two degrees of freedom coupled by a spring, each in contact with its own thermal bath. Under this mapping, the measurement and filtering process acts as an effective thermal bath at temperature $T_D=k\sigma^2/(2\tau_f)$, and the information flow can be interpreted as the entropy associated with the heat conduction between the particle and the filtered signal. Feedback cooling, $T>T_D$, then corresponds to heat conduction from the hotter physical bath through the particle to the colder effective bath, so that the mean medium entropy production of the particle is negative. This is allowed because the information flow, $\langle I^s_x\rangle$, which is negative as well, bounds the mean medium entropy production from below, according to Eq.~\eqref{eq:2ndlaw-x}. We further confirmed these results, as well as the fluctuation theorems, by numerical simulations.

As an outlook, a natural extension of the current problem is to combine the stochastic protocol with a prescribed one, for example, a time-dependent feedback stiffness. In this case, the dynamics becomes explicitly time-dependent, and the variations of both types of protocol contribute to the non-autonomous entropy production. Future work could also address the measurement, filtering, and feedback process, which has been treated here as a black box. An inclusive description of the measurement and feedback device, such as within the Hamiltonian framework of Ref.~\cite{DeffnerJarzynski2013}, would provide a more explicit interpretation of the measurement-and-filtering term $\Delta S_{\mathrm{m\&f}}$ and the information flow in terms of the physical degrees of freedom of the device.

\begin{acknowledgments}
This research was supported by the U.S. National Science Foundation under Grant No. DMR-2127900. The author thanks Christopher Jarzynski for guidance and Debankur Bhattacharyya for discussions.
\end{acknowledgments}

\onecolumngrid
\appendix
 
\section{Path-integral representation of overdamped Langevin dynamics}
\label{app:path}
 
In this appendix, we review the path-integral representation of overdamped Langevin dynamics~\cite{OnsagerMachlup1953,Chernyak2006} that is used throughout the main text. The fluctuation theorems in Secs.~\ref{sec:total} and \ref{sec:subsystem} are derived by comparing the path probability of the original dynamics, Eqs.~\eqref{eq:EOM}, with that of the reverse process or of a reference dynamics, Eqs.~\eqref{eq:adj-x} and \eqref{eq:adj-D}. All of these dynamics share the same diffusion coefficients and differ only in their drift terms. We therefore consider the general bipartite overdamped dynamics
\begin{equation}
  \dot q_i = f_i(\bm q,\lambda) + \sqrt{2\Di}\,\xi_i(t) \quad,\quad i\in\{x,D\},
  \label{app:eq:general}
\end{equation}
where $\bm q=(x,D)$, with components $q_x=x$ and $q_D=D$. Here, $f_i$ is the drift term of $q_i$, $\Di$ is the corresponding diffusion coefficient, and $\xi_x(t)$ and $\xi_D(t)$ are independent Gaussian white noises satisfying Eq.~\eqref{eq:noise}. The control parameter $\lambda$ is included to cover the driven dynamics with a prescribed protocol $\lambda_t$, Eq.~\eqref{eq:driven}; for the original dynamics, $f_i=\mu_i$ does not depend on $\lambda$, and $\Di$ is given by Eq.~\eqref{eq:diffusion}.
 
\subsection{Discretization and transition probability}
\label{app:path-disc}
 
Although the dynamics is continuous in time, it is convenient to discretize time when constructing the path probability. We partition the time interval $[0,\T]$ into $M$ segments of duration $\Delta t=\T/M\ll 1$, with $t_n=n\Delta t$, and denote $\bm q_n=\bm q(t_n)$. The discretized trajectory is given by $\Path=\{\bm q_0,\bm q_1,\dots,\bm q_M\}$. Since the dynamics is Markovian, the conditional probability of observing $\Path$, given the initial state $\bm q_0$, factorizes into a product of transition probabilities,
\begin{equation}
  P[\Path|\bm q_0]\,\mathcal{D}[\Path] = \prod_{n=0}^{M-1} P(\bm q_{n+1}|\bm q_n)\,\dd x_{n+1}\,\dd D_{n+1} ,
  \label{app:eq:factorize}
\end{equation}
where $\mathcal{D}[\Path]=\prod_{n=0}^{M-1}\dd x_{n+1}\,\dd D_{n+1}$ is the path-integral measure.
 
Using the Stratonovich (midpoint) discretization rule, Eq.~\eqref{app:eq:general} takes the discrete form
\begin{equation}
  q_{i,n+1} = q_{i,n} + f_i(\bar{\bm q}_n,\lambda_{\bar t_n})\,\Delta t + \sqrt{2\Di}\,\Delta W_{i,n} ,
  \label{app:eq:disc}
\end{equation}
where $q_{i,n}$ denotes the $i$-th component of $\bm q_n$, $\bar{\bm q}_n=(\bm q_{n+1}+\bm q_n)/2$ is the midpoint between $\bm q_n$ and $\bm q_{n+1}$, and $\bar t_n=(t_n+t_{n+1})/2 = t_n+\Delta t/2$. The Wiener increments $\Delta W_{x,n}$ and $\Delta W_{D,n}$, which we collect into $\bm{\Delta W}_n=(\Delta W_{x,n},\Delta W_{D,n})$, are independent and Gaussian distributed with zero mean and variance $\Delta t$,
\begin{equation}
  P(\bm{\Delta W}_n) = \prod_{i\in\{x,D\}}\frac{1}{\sqrt{2\pi\Delta t}}\exp\Bigl(-\frac{(\Delta W_{i,n})^2}{2\Delta t}\Bigr) .
  \label{app:eq:wiener}
\end{equation}
Since the noise in Eq.~\eqref{app:eq:general} is additive, the It\^o and Stratonovich interpretations of the equation of motion coincide. We nonetheless adopt the midpoint rule for the path probability, as it is symmetric under time reversal, i.e., the midpoints of a forward trajectory are mapped onto the same midpoints of its time-reversed counterpart. Moreover, the ordinary chain rule holds in the Stratonovich convention.
 
Define the displacement of the $n$-th step, $\epsilon_{i,n}=q_{i,n+1}-q_{i,n}$. For a fixed $\bm q_n$, Eq.~\eqref{app:eq:disc} establishes a one-to-one mapping between $\bm q_{n+1}$ and $\bm{\Delta W}_n$,
\begin{equation}
  \Delta W_{i,n} = \frac{\Delta t}{\sqrt{2\Di}}\Bigl[\frac{\epsilon_{i,n}}{\Delta t} - f_i(\bar{\bm q}_n,\lambda_{\bar t_n})\Bigr] .
  \label{app:eq:map}
\end{equation}
By probability conservation under the change of variables, the transition probability is
\begin{equation}
  P(\bm q_{n+1}|\bm q_n) = P\bigl(\bm{\Delta W}_n\bigr)\,\Bigl|\det\frac{\partial\bm{\Delta W}_n}{\partial\bm q_{n+1}}\Bigr| ,
  \label{app:eq:change-var}
\end{equation}
with $\Delta W_{i,n}$ given by Eq.~\eqref{app:eq:map}. Since $\partial\bar q_{j,n}/\partial q_{j,n+1}=1/2$, directly evaluating the elements of the Jacobian matrix $\partial\bm{\Delta W}_n/\partial\bm q_{n+1}$ gives
\begin{equation}
  \frac{\partial\Delta W_{i,n}}{\partial q_{j,n+1}} = \frac{1}{\sqrt{2\Di}}\Bigl[\delta_{ij}-\frac{1}{2}\partial_jf_i(\bar{\bm q}_n,\lambda_{\bar t_n})\,\Delta t\Bigr] ,
\end{equation}
where $\partial_j=\partial/\partial q_j$. The off-diagonal elements are of order $\Delta t$ and contribute to the determinant only at order $\Delta t^2$. Hence, to leading order in $\Delta t$, the determinant is given by the product of the diagonal elements,
\begin{equation}
  \Bigl|\det\frac{\partial\bm{\Delta W}_n}{\partial\bm q_{n+1}}\Bigr| = \frac{1}{\sqrt{4\Dx\DD}}\Bigl[1-\frac{1}{2}\sum_{i\in\{x,D\}}\partial_if_i(\bar{\bm q}_n,\lambda_{\bar t_n})\,\Delta t\Bigr] + O(\Delta t^2) .
\end{equation}
For $\Delta t\ll1$, this can be written in the exponential form
\begin{equation}
  \Bigl|\det\frac{\partial\bm{\Delta W}_n}{\partial\bm q_{n+1}}\Bigr| = \frac{1}{\sqrt{4\Dx\DD}}\exp\Bigl(-\frac{1}{2}\sum_{i\in\{x,D\}}\partial_if_i(\bar{\bm q}_n,\lambda_{\bar t_n})\,\Delta t\Bigr) + O(\Delta t^2) .
  \label{app:eq:jacobian}
\end{equation}
Substituting Eqs.~\eqref{app:eq:wiener}, \eqref{app:eq:map}, and \eqref{app:eq:jacobian} into Eq.~\eqref{app:eq:change-var} yields the transition probability
\begin{equation}
  P(\bm q_{n+1}|\bm q_n) = \prod_{i\in\{x,D\}}\frac{1}{\sqrt{4\pi\Di\Delta t}}\exp\Bigl(-\frac{\Delta t}{4\Di}\Bigl[\frac{\epsilon_{i,n}}{\Delta t}-f_i(\bar{\bm q}_n,\lambda_{\bar t_n})\Bigr]^2-\frac{1}{2}\partial_if_i(\bar{\bm q}_n,\lambda_{\bar t_n})\,\Delta t\Bigr) + O(\Delta t^{3/2}) .
  \label{app:eq:transition}
\end{equation}
 
\subsection{Onsager-Machlup function and path probability}
\label{app:path-OM}

Inserting Eq.~\eqref{app:eq:transition} into Eq.~\eqref{app:eq:factorize} gives the discrete path probability. Taking the continuum limit $\Delta t\to0$, we obtain the conditional path probability
\begin{align}
  P[\Path|\bm q_0] &= \lim_{\Delta t\to0}\prod_{n=0}^{M-1}\prod_{i\in\{x,D\}}\frac{1}{\sqrt{4\pi\Di\Delta t}}\exp\Bigl(-\frac{\Delta t}{4\Di}\Bigl[\frac{\epsilon_{i,n}}{\Delta t}-f_i(\bar{\bm q}_n,\lambda_{\bar t_n})\Bigr]^2-\frac{1}{2}\partial_if_i(\bar{\bm q}_n,\lambda_{\bar t_n})\,\Delta t\Bigr) \nonumber\\
  &= \mathcal{N}\lim_{\Delta t\to0}\exp\Bigl\{-\sum_{n=0}^{M-1}\sum_{i\in\{x,D\}}\Bigl(\frac{1}{4\Di}\Bigl[\frac{\epsilon_{i,n}}{\Delta t}-f_i(\bar{\bm q}_n,\lambda_{\bar t_n})\Bigr]^2+\frac{1}{2}\partial_if_i(\bar{\bm q}_n,\lambda_{\bar t_n})\Bigr)\Delta t\Bigr\} \nonumber\\
  &= \mathcal{N}\exp\Bigl(-\int_0^\T \OM(\bm q_t,\dot{\bm q}_t;\lambda_t)\,\dd t\Bigr) ,
  \label{app:eq:cond-path}
\end{align}
where $\mathcal{N}=\lim_{\Delta t\to0}\prod_{n=0}^{M-1}\prod_{i}(4\pi\Di\Delta t)^{-1/2}$ is the normalization factor, and the Onsager-Machlup function~\cite{OnsagerMachlup1953} is the sum of the contributions from the two subsystems,
\begin{equation}
  \OM = \sum_{i\in\{x,D\}}\OM_i[f_i] \quad,\quad
  \OM_i[f_i](\bm q,\dot q_i;\lambda) = \frac{1}{4\Di}\bigl(\dot q_i-f_i(\bm q,\lambda)\bigr)^2 + \frac{1}{2}\partial_if_i(\bm q,\lambda) .
  \label{app:eq:OM}
\end{equation}
Here, the notation $\OM_i[f_i]$ indicates that the contribution of subsystem $i$ depends on its drift $f_i$. The quadratic terms in Eq.~\eqref{app:eq:OM} are Stratonovich products and should therefore be evaluated according to the midpoint rule defined above. Since $\mathcal{N}$ depends only on the diffusion coefficients, it is identical for all dynamics considered in this paper and cancels in every ratio of path probabilities. For the original dynamics, $f_i=\mu_i$, substituting Eqs.~\eqref{eq:mobility} and~\eqref{eq:diffusion} into Eq.~\eqref{app:eq:OM} recovers Eq.~\eqref{eq:OM}.

The probability of observing the trajectory $\Path$ is obtained by weighting the conditional path probability with the initial distribution $\rho_0(\bm q_0)$,
\begin{equation}
  P[\Path] = P[\Path|\bm q_0]\,\rho_0(\bm q_0) ,
  \label{app:eq:path}
\end{equation}
and the ensemble average of an arbitrary trajectory-dependent observable $A[\Path]$ can be written as a path integral over all possible trajectories,
\begin{equation}
  \langle A\rangle=\int\mathcal{D}[\Path]\,A[\Path]\,P[\Path] .
  \label{app:eq:avg}
\end{equation}
In the following appendixes, we use this representation to compare the probability of a trajectory with that of its time-reversed counterpart under the reverse process or a reference dynamics.

\section{Fluctuation theorems for the total entropy production}
\label{app:total}

In this appendix, we derive the DFT for the total entropy production at the trajectory level, Eqs.~\eqref{eq:DFT-traj}--\eqref{eq:Stot-terms}, and promote it to the DFT for the probability distribution of $\Stot$ and the corresponding IFT, Eq.~\eqref{eq:DFT-tot}.

\subsection{Detailed fluctuation theorem at the trajectory level}
\label{app:total-traj}

Using Eqs.~\eqref{eq:pathprob} and \eqref{eq:pathprob-rev-changed}, the ratio of the probability of observing the trajectory $\Path$ to that of observing its time-reversed counterpart $\Path^\dagger$ is
\begin{align}
  \frac{P[\Path]}{P[\Path^\dagger]} &= \frac{P[\Path|x_0,D_0]}{P[\Path^\dagger|x_\T,D_\T]}\,\frac{\rs(x_0,D_0)}{\rs(x_\T,D_\T)} \nonumber\\
  &= \exp\Bigl(-\int_0^\T\Bigl[\mathbf{S}(x_t,\dot x_t,D_t,\dot D_t)-\mathbf{S}(x_t,-\dot x_t,D_t,-\dot D_t)\Bigr]\dd t + \bigl[\phi(x_\T,D_\T)-\phi(x_0,D_0)\bigr]\Bigr) ,
  \label{app:eq:tot-ratio}
\end{align}
where the normalization factors cancel. Substituting the explicit form of the Onsager-Machlup function, Eq.~\eqref{eq:OM}, the difference of the Onsager-Machlup functions gives
\begin{equation}
  \mathbf{S}(x,\dot x,D,\dot D)-\mathbf{S}(x,-\dot x,D,-\dot D) = -\frac{1}{T}\,\dot x\strat F - \frac{2\tau_f}{\sigma^2}\,\dot D\strat\bigl[O(x)-D\bigr] .
\end{equation}
Substituting this back into Eq.~\eqref{app:eq:tot-ratio}, we obtain
\begin{equation}
  \frac{P[\Path]}{P[\Path^\dagger]} = \exp\Bigl(\frac{1}{T}\int_0^\T\dot x_t\strat F(x_t,D_t)\,\dd t + \frac{2\tau_f}{\sigma^2}\int_0^\T\dot D_t\strat\bigl[O(x_t)-D_t\bigr]\dd t + \bigl[\phi(x_\T,D_\T)-\phi(x_0,D_0)\bigr]\Bigr) .
  \label{app:eq:tot-ratio-explicit}
\end{equation}
We now identify the three terms in the exponent. From stochastic energetics~\cite{Sekimoto1997,Sekimoto2010}, the heat absorbed by the particle from the thermal bath is
\begin{equation}
  Q[\Path] = \int_0^\T\dot x_t\strat\partial_x\bigl[U(x_t)+V(x_t,D_t)\bigr]\,\dd t = -\int_0^\T\dot x_t\strat F(x_t,D_t)\,\dd t .
\end{equation}
Hence, the first term in the exponent can be rewritten as
\begin{equation}
  \Smed[\Path] = \frac{1}{T}\int_0^\T\dot x_t\strat F(x_t,D_t)\,\dd t = -\frac{Q[\Path]}{T} ,
\end{equation}
which is the entropy change in the thermal bath due to the heat exchange with the particle, i.e., the medium entropy production. The second term arises from the stochastic dynamics of $D$ generated by the measurement and filtering process. We therefore define
\begin{equation}
  \Smf[\Path] = \frac{2\tau_f}{\sigma^2}\int_0^\T\dot D_t\strat\bigl[O(x_t)-D_t\bigr]\dd t
\end{equation}
as the entropy production associated with the measurement and filtering process. Note that $\Smf$ does not depend on the feedback potential $V(x,D)$ and persists in the absence of feedback, i.e., for $V=0$. This term therefore does not originate from the feedback. The third term depends only on the initial and final states of the trajectory. Since $\phi=-\ln\rs$ is the stochastic Shannon entropy of the joint system in the stationary state~\cite{Seifert2005}, it is the change in the system entropy along the trajectory,
\begin{equation}
  \Ssys[\Path] = \phi(x_\T,D_\T)-\phi(x_0,D_0) .
\end{equation}
These three terms are reported in Eqs.~\eqref{eq:Smed}--\eqref{eq:Ssys}. Defining the total entropy production along the trajectory $\Path$ as their sum,
\begin{equation}
  \Stot[\Path] = \Smed[\Path] + \Smf[\Path] + \Ssys[\Path] ,
\end{equation}
Eq.~\eqref{app:eq:tot-ratio-explicit} becomes
\begin{equation}
  \frac{P[\Path]}{P[\Path^\dagger]} = \exp\bigl(\Smed[\Path]+\Smf[\Path]+\Ssys[\Path]\bigr) = \ee^{\Stot[\Path]} ,
\end{equation}
which is the DFT for the total entropy production at the trajectory level, Eq.~\eqref{eq:DFT-traj}.

To proceed, we note that the total entropy production is odd under time reversal. By explicit evaluation,
\begin{align}
  \Stot[\Path^\dagger] &= \frac{1}{T}\int_0^\T\dot x^\dagger_t\strat F(x^\dagger_t,D^\dagger_t)\,\dd t + \frac{2\tau_f}{\sigma^2}\int_0^\T\dot D^\dagger_t\strat\bigl[O(x^\dagger_t)-D^\dagger_t\bigr]\dd t + \bigl[\phi(x^\dagger_\T,D^\dagger_\T)-\phi(x^\dagger_0,D^\dagger_0)\bigr] \nonumber\\
  &= -\frac{1}{T}\int_0^\T\dot x_t\strat F(x_t,D_t)\,\dd t - \frac{2\tau_f}{\sigma^2}\int_0^\T\dot D_t\strat\bigl[O(x_t)-D_t\bigr]\dd t + \bigl[\phi(x_0,D_0)-\phi(x_\T,D_\T)\bigr] \nonumber\\
  &= -\Stot[\Path] ,
  \label{app:eq:Stot-odd}
\end{align}
where, from the first to the second line, we performed the change of integration variable $t\to\T-t$, under which $\dot x^\dagger_t\to-\dot x_t$ and $\dot D^\dagger_t\to-\dot D_t$.

\subsection{Detailed fluctuation theorem for the distribution and integral fluctuation theorem}
\label{app:general}

Let $p(\Stot)$ denote the probability distribution of the total entropy production. It can be expressed as a path integral over all trajectories $\Path$ satisfying $\Stot[\Path]=\Stot$,
\begin{equation}
  p(\Stot) = \int\mathcal{D}[\Path]\,P[\Path]\,\delta\bigl(\Stot-\Stot[\Path]\bigr) .
  \label{app:eq:dist}
\end{equation}
Using Eqs.~\eqref{eq:DFT-traj} and \eqref{app:eq:Stot-odd}, together with a change of variables $\mathcal{D}[\Path]\to\mathcal{D}[\Path^\dagger]$, we have
\begin{align}
  p(\Stot) &= \int\mathcal{D}[\Path]\,P[\Path^\dagger]\,\ee^{\Stot[\Path]}\,\delta\bigl(\Stot-\Stot[\Path]\bigr) \nonumber\\
  &= \ee^{\Stot}\int\mathcal{D}[\Path]\,P[\Path^\dagger]\,\delta\bigl(\Stot-\Stot[\Path]\bigr) \nonumber\\
  &= \ee^{\Stot}\int\mathcal{D}[\Path^\dagger]\,P[\Path^\dagger]\,\delta\bigl(\Stot+\Stot[\Path^\dagger]\bigr) \nonumber\\
  &= \ee^{\Stot}\,p(-\Stot) .
  \label{app:eq:dist-steps}
\end{align}
Here, time reversal only relabels the points of the trajectory, $\bm q_n\to\bm q_{M-n}$, so the change of variables has unit Jacobian and the path measure remains unchanged, $\mathcal{D}[\Path]=\mathcal{D}[\Path^\dagger]$. Since the reverse process obeys the same dynamics with the same initial distribution $\rs$, the path integral in the third line yields $p(-\Stot)$. We thus arrive at the DFT for the probability distribution of the total entropy production,
\begin{equation}
  \frac{p(\Stot)}{p(-\Stot)} = \ee^{\Stot} .
  \label{app:eq:DFT-dist}
\end{equation}

The corresponding IFT follows directly from Eq.~\eqref{app:eq:DFT-dist}. Rewriting it as
\begin{equation}
  p(\Stot)\,\ee^{-\Stot} = p(-\Stot)
\end{equation}
and integrating both sides over $\Stot$, the left-hand side yields the ensemble average $\langle \ee^{-\Stot}\rangle$, while the right-hand side evaluates to unity due to normalization. Therefore, we obtain the IFT
\begin{equation}
  \langle \ee^{-\Stot}\rangle = 1 .
  \label{app:eq:IFT}
\end{equation}
Finally, applying Jensen's inequality, $\langle \ee^{-\Stot}\rangle\ge\ee^{-\langle \Stot\rangle}$, to Eq.~\eqref{app:eq:IFT} gives the second law, $\langle \Stot\rangle\ge0$.

\section{Fluctuation theorems for non-autonomous and housekeeping entropy production}
\label{app:prescribed}

In this appendix, we derive the fluctuation theorem for the non-autonomous entropy production, Eqs.~\eqref{eq:FT-na} and \eqref{eq:na}, together with that for the housekeeping entropy production, for the driven dynamics with a prescribed protocol, Eq.~\eqref{eq:driven}~\cite{HatanoSasa2001,Chernyak2006,EspositoVdB2010PRL,VdBEsposito2010PRE}. We assume that, for every fixed value of $\lambda$, the dynamics admits a stationary distribution $\rs(x;\lambda)$, with the stationary probability current
\begin{equation}
  J^s(x;\lambda) = f(x;\lambda)\rs(x;\lambda) - \mathcal{D}\partial_x\rs(x;\lambda)
  \label{app:eq:pres-current}
\end{equation}
satisfying $\partial_xJ^s=0$. The Onsager-Machlup function of the dynamics follows from Eq.~\eqref{app:eq:OM} with a single degree of freedom,
\begin{equation}
  \mathbf{S}(x,\dot x;\lambda) = \frac{1}{4\mathcal{D}}\bigl(\dot x - f(x;\lambda)\bigr)^2 + \frac{1}{2}\partial_xf(x;\lambda) .
  \label{app:eq:pres-OM}
\end{equation}

\subsection{Housekeeping and non-autonomous entropy production}
\label{app:pres-decomp}

Let $\mathbf{X}=\{x_t\,|\,t\in[0,\T]\}$ denote a trajectory of the system as $\lambda$ varies according to the protocol $\lambda_t$, with the initial condition sampled from $\rs(x;\lambda_0)$; this defines the forward process. Following the same steps as in Appendix~\ref{app:total}, the total entropy production along $\mathbf{X}$ is
\begin{equation}
  \Stot[\mathbf{X}] = \int_0^\T\dot x_t\strat\frac{f(x_t;\lambda_t)}{\mathcal{D}}\,\dd t - \bigl[\ln\rs(x_\T;\lambda_\T)-\ln\rs(x_0;\lambda_0)\bigr] .
\end{equation}
Since the term in the square brackets is evaluated only at the initial and final times, it can be written as the time integral of a total time derivative. Applying the chain rule, we obtain
\begin{equation}
  \Stot[\mathbf{X}] = \int_0^\T\Bigl[\dot x_t\strat\frac{f(x_t;\lambda_t)}{\mathcal{D}} - \dot x_t\strat\partial_x\ln\rs(x_t;\lambda_t) - \dot\lambda_t\,\partial_\lambda\ln\rs(x_t;\lambda_t)\Bigr]\dd t .
\end{equation}
Accordingly, we define the housekeeping entropy production as the sum of the first two terms, which are associated with the dynamics of the system, and the non-autonomous entropy production as the last term, which arises from the time dependence of the control parameter,
\begin{equation}
  \Delta S_{\mathrm{hk}}[\mathbf{X}] = \int_0^\T\dot x_t\strat\Bigl(\frac{f(x_t;\lambda_t)}{\mathcal{D}}-\partial_x\ln\rs(x_t;\lambda_t)\Bigr)\dd t \quad,\quad
  \Delta S_{\mathrm{na}}[\mathbf{X}] = -\int_0^\T\dot\lambda_t\,\partial_\lambda\ln\rs(x_t;\lambda_t)\,\dd t .
  \label{app:eq:hk-na}
\end{equation}
By construction, the total entropy production thus admits the decomposition
\begin{equation}
    \Stot=\Delta S_{\mathrm{hk}}+\Delta S_{\mathrm{na}} .
\end{equation}
The non-autonomous entropy production vanishes if $\lambda$ is held fixed. It also vanishes in the quasistatic limit, $\dot\lambda\to0$. On the other hand, the housekeeping entropy production vanishes in the absence of non-conservative forces. For a conservative force, $f=-\partial_xU/\gamma$, the Einstein relation $\mathcal{D}=T/\gamma$ and the stationary distribution $\rs\propto\ee^{-U/T}$ give $f/\mathcal{D}=-\partial_xU/T=\partial_x\ln\rs$, so that $\Delta S_{\mathrm{hk}}$ vanishes identically by Eq.~\eqref{app:eq:hk-na}.

\subsection{Adjoint dynamics}
\label{app:pres-adj}

The adjoint dynamics is defined as the dynamics that relaxes to the same stationary distribution $\rs(x;\lambda)$ as the original dynamics, but with the opposite stationary probability current~\cite{Chernyak2006,EspositoVdB2010PRL,VdBEsposito2010PRE}. It is governed by the equation of motion
\begin{equation}
  \dot x = \hat f(x;\lambda) + \sqrt{2\mathcal{D}}\,\xi(t) \quad,\quad \hat f(x;\lambda) = -f(x;\lambda) + 2\mathcal{D}\partial_x\ln\rs(x;\lambda) .
  \label{app:eq:pres-adj}
\end{equation}
Substituting $\rs$ into the probability current of the adjoint dynamics gives
\begin{equation}
  \hat J^s = \hat f\rs - \mathcal{D}\partial_x\rs = -f\rs + 2\mathcal{D}\partial_x\rs - \mathcal{D}\partial_x\rs = -J^s .
\end{equation}
Hence, $-\partial_x\hat J^s=\partial_xJ^s=0$, which confirms that $\rs(x;\lambda)$ is also the stationary distribution of the adjoint dynamics, with the opposite stationary current. The corresponding Onsager-Machlup function is $\hat{\mathbf{S}}(x,\dot x;\lambda)$, given by Eq.~\eqref{app:eq:pres-OM} with $f$ replaced by $\hat f$.

\subsection{Fluctuation theorem for the housekeeping entropy production}
\label{app:pres-hk}

To derive the fluctuation theorem for the housekeeping entropy production, we choose the forward trajectory $\mathbf{X}$ as the reference trajectory and the adjoint dynamics with the forward protocol $\lambda_t$ as the reference dynamics, with the initial condition also sampled from $\rs(x;\lambda_0)$. Since the initial distributions are the same for both the forward and the reference dynamics, the ratio of the path probabilities gives
\begin{equation}
  \frac{P^F[\mathbf{X}]}{\hat P^F[\mathbf{X}]} = \exp\Bigl(-\int_0^\T\Bigl[\mathbf{S}(x_t,\dot x_t;\lambda_t)-\hat{\mathbf{S}}(x_t,\dot x_t;\lambda_t)\Bigr]\dd t\Bigr) .
  \label{app:eq:hk-ratio}
\end{equation}
Using Eqs.~\eqref{app:eq:pres-OM} and \eqref{app:eq:pres-adj}, the difference of the Onsager-Machlup functions gives
\begin{equation}
  \mathbf{S}(x,\dot x;\lambda)-\hat{\mathbf{S}}(x,\dot x;\lambda) = -\dot x\strat\Bigl(\frac{f}{\mathcal{D}}-\partial_x\ln\rs\Bigr) + \frac{1}{\rs}\Bigl[\partial_x\bigl(f\rs\bigr)-\mathcal{D}\partial_x^2\rs\Bigr] .
\end{equation}
The term in the square brackets equals $\partial_xJ^s$ and vanishes by stationarity. Substituting this result back into Eq.~\eqref{app:eq:hk-ratio}, we obtain the DFT for the housekeeping entropy production at the trajectory level,
\begin{equation}
  \frac{P^F[\mathbf{X}]}{\hat P^F[\mathbf{X}]} = \ee^{\Delta S_{\mathrm{hk}}[\mathbf{X}]} .
  \label{app:eq:FT-hk}
\end{equation}

\subsection{Fluctuation theorem for the non-autonomous entropy production}
\label{app:pres-na}

To derive the fluctuation theorem for the non-autonomous entropy production, we choose the time-reversed trajectory $\mathbf{X}^\dagger=\{x_{\T-t}\,|\,t\in[0,\T]\}$ as the reference trajectory, and the adjoint dynamics with the reversed protocol $\lambda^R_t=\lambda_{\T-t}$ as the reference dynamics. Assuming that the initial condition is sampled from $\rs(x;\lambda^R_0)=\rs(x;\lambda_\T)$, the probability of observing $\mathbf{X}^\dagger$ under this reverse adjoint process is
\begin{align}
  \hat P^R[\mathbf{X}^\dagger] &= \mathcal{N}\exp\Bigl(-\int_0^\T\hat{\mathbf{S}}(x^\dagger_t,\dot x^\dagger_t;\lambda^R_t)\,\dd t\Bigr)\rs(x_\T;\lambda_\T) \nonumber\\
  &= \mathcal{N}\exp\Bigl(-\int_0^\T\hat{\mathbf{S}}(x_t,-\dot x_t;\lambda_t)\,\dd t\Bigr)\rs(x_\T;\lambda_\T) .
\end{align}
Here, going from the first to the second line, we performed the change of integration variable $t\to\T-t$. Comparing with the probability of observing $\mathbf{X}$ under the forward process, we have
\begin{equation}
  \frac{P^F[\mathbf{X}]}{\hat P^R[\mathbf{X}^\dagger]} = \exp\Bigl(-\int_0^\T\Bigl[\mathbf{S}(x_t,\dot x_t;\lambda_t)-\hat{\mathbf{S}}(x_t,-\dot x_t;\lambda_t)\Bigr]\dd t - \bigl[\ln\rs(x_\T;\lambda_\T)-\ln\rs(x_0;\lambda_0)\bigr]\Bigr) .
  \label{app:eq:na-ratio}
\end{equation}
Using Eqs.~\eqref{app:eq:pres-OM} and \eqref{app:eq:pres-adj}, the difference of the Onsager-Machlup functions gives
\begin{equation}
  \mathbf{S}(x,\dot x;\lambda)-\hat{\mathbf{S}}(x,-\dot x;\lambda) = -\dot x\strat\partial_x\ln\rs + \frac{1}{\rs}\Bigl[\partial_x\bigl(f\rs\bigr)-\mathcal{D}\partial_x^2\rs\Bigr] = -\dot x\strat\partial_x\ln\rs ,
  \label{app:eq:diff-OM-na}
\end{equation}
where the term in the square brackets again vanishes by stationarity. Writing the boundary term in Eq.~\eqref{app:eq:na-ratio} as the time integral of a total time derivative and applying the chain rule, we have
\begin{equation}
  \ln\rs(x_\T;\lambda_\T)-\ln\rs(x_0;\lambda_0) = \int_0^\T\Bigl[\dot x_t\strat\partial_x\ln\rs(x_t;\lambda_t) + \dot\lambda_t\,\partial_\lambda\ln\rs(x_t;\lambda_t)\Bigr]\dd t .
  \label{app:eq:SE-na}
\end{equation}
Substituting Eqs.~\eqref{app:eq:diff-OM-na} and \eqref{app:eq:SE-na} into Eq.~\eqref{app:eq:na-ratio}, the terms proportional to $\dot x_t$ cancel, and we arrive at
\begin{equation}
  \frac{P^F[\mathbf{X}]}{\hat P^R[\mathbf{X}^\dagger]} = \exp\Bigl(-\int_0^\T\dot\lambda_t\,\partial_\lambda\ln\rs(x_t;\lambda_t)\,\dd t\Bigr) = \ee^{\Delta S_{\mathrm{na}}[\mathbf{X}]} ,
  \label{app:eq:FT-na}
\end{equation}
which is reported in Eqs.~\eqref{eq:FT-na} and \eqref{eq:na} of the main text.

From the trajectory-level DFTs, Eqs.~\eqref{app:eq:FT-hk} and \eqref{app:eq:FT-na}, following the analogous steps as in Appendix~\ref{app:general}, one obtains the DFTs for the probability distributions and the corresponding IFTs,
\begin{equation}
  \frac{p(\Delta S_{\mathrm{hk}})}{\hat p^F(-\Delta S_{\mathrm{hk}})} = \ee^{\Delta S_{\mathrm{hk}}} \quad,\quad \langle \ee^{-\Delta S_{\mathrm{hk}}}\rangle = 1 \quad,\quad
  \frac{p(\Delta S_{\mathrm{na}})}{\hat p^R(-\Delta S_{\mathrm{na}})} = \ee^{\Delta S_{\mathrm{na}}} \quad,\quad \langle \ee^{-\Delta S_{\mathrm{na}}}\rangle = 1 ,
\end{equation}
where $\hat p^F$ and $\hat p^R$ denote the probability distributions of $-\Delta S_{\mathrm{hk}}[\mathbf{X}]$ and $-\Delta S_{\mathrm{na}}[\mathbf{X}^\dagger]$, respectively, for trajectories $\mathbf{X}$ generated by the forward and reverse adjoint processes. 

\section{Fluctuation theorems for the subsystem entropy production}
\label{app:sub}

In this appendix, we derive the trajectory-level DFTs for the subsystem entropy production, Eqs.~\eqref{eq:FT-D} and \eqref{eq:FT-x}, and the corresponding DFTs for the probability distributions and IFTs, Eqs.~\eqref{eq:IFT-sub}. Throughout this appendix, $\rs(x,D)$ and $J^s_i$ denote the stationary distribution and the stationary partial currents of the original dynamics, Eqs.~\eqref{eq:EOM}, which satisfy
\begin{equation}
  \partial_x\Jx = -\partial_D\JD .
  \label{app:eq:sub-stat}
\end{equation}
By Eq.~\eqref{app:eq:OM}, the Onsager-Machlup function of the original dynamics, Eq.~\eqref{eq:OM}, is the sum of the contributions from the two subsystems, $\mathbf{S}=\OM_x[\mu_x]+\OM_D[\mu_D]$, where
\begin{equation}
  \OM_x[\mu_x](\bm q,\dot x) = \frac{1}{4\Dx}\bigl(\dot x-\mu_x\bigr)^2 + \frac{1}{2}\partial_x\mu_x \quad,\quad
  \OM_D[\mu_D](\bm q,\dot D) = \frac{1}{4\DD}\bigl(\dot D-\mu_D\bigr)^2 + \frac{1}{2}\partial_D\mu_D .
  \label{app:eq:sub-OM}
\end{equation}

\subsection{Reference dynamics}
\label{app:sub-ref}

The reference dynamics for $\Delta S_D$, Eq.~\eqref{eq:adj-x}, is obtained by replacing the drift of $x$ with its adjoint, $\hat\mu_x=-\mu_x+2\Dx\partial_x\ln\rs$, while leaving the drift of $D$ unchanged. Its Onsager-Machlup function is therefore $\hat{\mathbf{S}}_x=\OM_x[\hat\mu_x]+\OM_D[\mu_D]$. Substituting $\rs$ into the partial currents of the reference dynamics gives
\begin{equation}
  \hat J^s_x = \hat\mu_x\rs-\Dx\partial_x\rs = -\Jx \quad,\quad J^s_D = \mu_D\rs-\DD\partial_D\rs .
\end{equation}
The rate of change of $\rs$ under the reference dynamics is thus
\begin{equation}
  -\partial_x\hat J^s_x-\partial_D\JD = \partial_x\Jx-\partial_D\JD = -2\,\partial_D\JD ,
  \label{app:eq:sub-nonstat}
\end{equation}
where we used Eq.~\eqref{app:eq:sub-stat}. Hence, unlike the adjoint dynamics for a prescribed protocol (Appendix~\ref{app:pres-adj}), $\rs$ is not the stationary distribution of the reference dynamics unless $\partial_D\JD=0$. This, however, does not affect the derivation of the fluctuation theorems below. As in Appendix~\ref{app:general}, the derivation uses only the normalization of the reference path probability, which holds as long as the reference dynamics is a well-defined Markov process and its initial distribution $\rs$ is normalized. Similarly, the reference dynamics for $\Delta S_x$, Eq.~\eqref{eq:adj-D}, has the Onsager-Machlup function $\hat{\mathbf{S}}_D=\OM_x[\mu_x]+\OM_D[\hat\mu_D]$, with $\hat\mu_D=-\mu_D+2\DD\partial_D\ln\rs$, and $\rs$ is not its stationary distribution unless $\partial_x\Jx=0$.

\subsection{Fluctuation theorem for $\Delta S_D$}
\label{app:sub-D}

We choose the time-reversed trajectory $\Path^\dagger$ as the reference trajectory, and the dynamics defined by Eq.~\eqref{eq:adj-x} as the reference dynamics. Assuming that the initial condition is sampled from $\rs$, the probability of observing $\Path^\dagger$ under the reference dynamics is
\begin{align}
  \hat P_x[\Path^\dagger] &= \mathcal{N}\exp\Bigl(-\int_0^\T\Bigl\{\OM_x[\hat\mu_x](\bm q^\dagger_t,\dot x^\dagger_t)+\OM_D[\mu_D](\bm q^\dagger_t,\dot D^\dagger_t)\Bigr\}\dd t\Bigr)\rs(x_\T,D_\T) \nonumber\\
  &= \mathcal{N}\exp\Bigl(-\int_0^\T\Bigl\{\OM_x[\hat\mu_x](\bm q_t,-\dot x_t)+\OM_D[\mu_D](\bm q_t,-\dot D_t)\Bigr\}\dd t\Bigr)\rs(x_\T,D_\T) .
\end{align}
Here, going from the first to the second line, we performed the change of integration variable $t\to\T-t$. Comparing with the probability of observing $\Path$ under the original dynamics, Eq.~\eqref{eq:pathprob}, we have
\begin{align}
  \frac{P[\Path]}{\hat P_x[\Path^\dagger]} &= \exp\Bigl(-\int_0^\T\Bigl[\mathbf{S}(\bm q_t,\dot{\bm q}_t)-\hat{\mathbf{S}}_x(\bm q_t,-\dot{\bm q}_t)\Bigr]\dd t - \bigl[\ln\rs(x_\T,D_\T)-\ln\rs(x_0,D_0)\bigr]\Bigr) \nonumber\\
  &= \exp\Bigl(-\int_0^\T\Bigl\{\OM_x[\mu_x](\bm q_t,\dot x_t)-\OM_x[\hat\mu_x](\bm q_t,-\dot x_t)\Bigr\}\dd t \nonumber\\
  &\qquad\quad -\int_0^\T\Bigl\{\OM_D[\mu_D](\bm q_t,\dot D_t)-\OM_D[\mu_D](\bm q_t,-\dot D_t)\Bigr\}\dd t - \bigl[\ln\rs(x_\T,D_\T)-\ln\rs(x_0,D_0)\bigr]\Bigr) .
  \label{app:eq:sub-D-ratio}
\end{align}
For the difference associated with the dynamics of $x$, the same calculation as in Eq.~\eqref{app:eq:diff-OM-na}, now with $\partial_x$ taken at fixed $D$, gives
\begin{equation}
  \OM_x[\mu_x](\bm q,\dot x)-\OM_x[\hat\mu_x](\bm q,-\dot x) = -\dot x\strat\partial_x\ln\rs + \frac{1}{\rs}\Bigl[\partial_x\bigl(\mu_x\rs\bigr)-\Dx\partial_x^2\rs\Bigr] = -\dot x\strat\partial_x\ln\rs + \frac{\partial_x\Jx}{\rs} .
  \label{app:eq:sub-D-diffx}
\end{equation}
In contrast to Appendix~\ref{app:pres-na}, the term $\partial_x\Jx$ does not vanish in general, since only the sum $\partial_x\Jx+\partial_D\JD$ vanishes in the stationary state. For the difference associated with the dynamics of $D$, substituting Eq.~\eqref{app:eq:sub-OM} yields
\begin{equation}
  \OM_D[\mu_D](\bm q,\dot D)-\OM_D[\mu_D](\bm q,-\dot D) =  -\dot D\strat\frac{\mu_D}{\DD} .
  \label{app:eq:sub-D-diffD}
\end{equation}
Writing the boundary term in Eq.~\eqref{app:eq:sub-D-ratio} as the time integral of a total time derivative and applying the chain rule, we have
\begin{equation}
  \ln\rs(x_\T,D_\T)-\ln\rs(x_0,D_0) = \int_0^\T\Bigl[\dot x_t\strat\partial_x\ln\rs(x_t,D_t) + \dot D_t\strat\partial_D\ln\rs(x_t,D_t)\Bigr]\dd t .
  \label{app:eq:sub-SE}
\end{equation}
Substituting Eqs.~\eqref{app:eq:sub-D-diffx}--\eqref{app:eq:sub-SE} into Eq.~\eqref{app:eq:sub-D-ratio}, the terms $\dot x\strat\partial_x\ln\rs$ cancel, and we obtain
\begin{equation}
  \frac{P[\Path]}{\hat P_x[\Path^\dagger]} = \exp\Bigl(\int_0^\T\Bigl[\dot D_t\strat\Bigl(\frac{\mu_D}{\DD}-\partial_D\ln\rs\Bigr)-\frac{\partial_x\Jx}{\rs}\Bigr]\dd t\Bigr) .
\end{equation}
Finally, with $-\partial_x\Jx=\partial_D\JD$ and $\mu_D/\DD=2\tau_f[O(x)-D]/\sigma^2$, which follow from Eq.~\eqref{app:eq:sub-stat} and Eqs.~\eqref{eq:mobility} and \eqref{eq:diffusion}, respectively, we arrive at
\begin{equation}
  \frac{P[\Path]}{\hat P_x[\Path^\dagger]} = \exp\Bigl(\int_0^\T\Bigl[\dot D_t\strat\Bigl(\frac{2\tau_f(O-D)}{\sigma^2}-\partial_D\ln\rs\Bigr)+\frac{\partial_D\JD}{\rs}\Bigr]\dd t\Bigr) = \ee^{\Delta S_D[\Path]} , \label{app:eq:FT-D}
\end{equation}
which is Eq.~\eqref{eq:FT-D} of the main text, with $\Delta S_D$ given by Eq.~\eqref{eq:SDpm}.

\subsection{Fluctuation theorem for $\Delta S_x$}
\label{app:sub-x}

The derivation for $\Delta S_x$ follows by exchanging the roles of $x$ and $D$ in Appendix~\ref{app:sub-D}. We compare $P[\Path]$ with the probability $\hat P_D[\Path^\dagger]$ of observing $\Path^\dagger$ under the reference dynamics defined by Eq.~\eqref{eq:adj-D}, again with the initial condition sampled from $\rs$. In this case, the dynamics of $x$ is compared with itself, and the dynamics of $D$ is compared with the time-reversed adjoint dynamics, which give
\begin{equation}
  \OM_x[\mu_x](\bm q,\dot x)-\OM_x[\mu_x](\bm q,-\dot x) = -\dot x\strat\frac{\mu_x}{\Dx} \quad,\quad
  \OM_D[\mu_D](\bm q,\dot D)-\OM_D[\hat\mu_D](\bm q,-\dot D) = -\dot D\strat\partial_D\ln\rs + \frac{\partial_D\JD}{\rs} .
\end{equation}
Substituting these differences and Eq.~\eqref{app:eq:sub-SE} into the ratio of the path probabilities, together with $-\partial_D\JD=\partial_x\Jx$ and $\mu_x/\Dx=F/T$, we obtain
\begin{align}
  \frac{P[\Path]}{\hat P_D[\Path^\dagger]} &= \exp\Bigl(-\int_0^\T\Bigl[\mathbf{S}(\bm q_t,\dot{\bm q}_t)-\hat{\mathbf{S}}_D(\bm q_t,-\dot{\bm q}_t)\Bigr]\dd t - \bigl[\ln\rs(x_\T,D_\T)-\ln\rs(x_0,D_0)\bigr]\Bigr) \nonumber\\
  &= \exp\Bigl(\int_0^\T\Bigl[\dot x_t\strat\Bigl(\frac{F}{T}-\partial_x\ln\rs\Bigr)+\frac{\partial_x\Jx}{\rs}\Bigr]\dd t\Bigr) = \ee^{\Delta S_x[\Path]} , \label{app:eq:FT-x}
\end{align}
which is Eq.~\eqref{eq:FT-x} of the main text, with $\Delta S_x$ given by Eq.~\eqref{eq:Sxpm}.
\subsection{Distributions and integral fluctuation theorems}
\label{app:sub-dist}

From the trajectory-level DFTs, Eqs.~\eqref{app:eq:FT-D} and \eqref{app:eq:FT-x}, following the analogous steps as in Appendix~\ref{app:general}, with $P[\Path^\dagger]$ replaced by $\hat P_x[\Path^\dagger]$ and $\hat P_D[\Path^\dagger]$, respectively, one obtains the DFTs for the probability distributions, the IFTs, and the second-law-like inequalities,
\begin{subequations}
\label{app:eq:sub-dist}
\begin{align}
  \frac{p(\Delta S_D)}{\hat p_x(-\Delta S_D)} &= \ee^{\Delta S_D} \quad,\quad
  \langle \ee^{-\Delta S_D}\rangle = 1 \quad,\quad
  \langle \Delta S_D\rangle\ge 0 , \label{app:eq:sub-dist-D}\\
  \frac{p(\Delta S_x)}{\hat p_D(-\Delta S_x)} &= \ee^{\Delta S_x} \quad,\quad
  \langle \ee^{-\Delta S_x}\rangle = 1 \quad,\quad
  \langle \Delta S_x\rangle\ge 0 , \label{app:eq:sub-dist-x}
\end{align}
\end{subequations}
which are Eqs.~\eqref{eq:IFT-sub} of the main text. Here, $\hat p_x$ and $\hat p_D$ denote the probability distributions of $-\Delta S_D[\Path^\dagger]$ and $-\Delta S_x[\Path^\dagger]$, respectively, for trajectories $\Path$ generated by the corresponding reference dynamics. Unlike the case of the total entropy production, $-\Delta S_i[\Path^\dagger]\neq\Delta S_i[\Path]$, since the current term in Eq.~\eqref{eq:S-general} is even under time reversal [Eq.~\eqref{eq:S-general-minus}]. Hence, $\hat p_x$ and $\hat p_D$ are in general not the distributions of $\Delta S_D$ and $\Delta S_x$ in the reference dynamics.

\section{Harmonic feedback model: stationary solution and energetics}
\label{app:energetics}

In this appendix, we provide the details for the solvable model of Sec.~\ref{sec:harmonic}, with $U=0$, $O(x)=x$, and $V(x,D)=k(x-D)^2/2$. We define the relative coordinate $y=x-D$, and the equations of motion, Eqs.~\eqref{eq:EOMharm}, can be written as
\begin{subequations}
\label{app:eq:EOMharm}
\begin{align}
  \dot x &= -\frac{k}{\gamma}\,y + \sqrt{\frac{2T}{\gamma}}\,\xi_x(t) , \\
  \dot D &= \frac{1}{\tau_f}\,y + \frac{\sigma}{\tau_f}\,\xi_D(t) .
\end{align}
\end{subequations}
We first verify the stationary solution, Eq.~\eqref{eq:betastar}, together with Eqs.~\eqref{eq:harm-derivs}, and then derive the explicit expressions for the heat and work, Eqs.~\eqref{eq:QW}, and their stationary means, Eq.~\eqref{eq:Qmean}.

\subsection{Stationary solution}
\label{app:harm-stat}

We consider a solution of the Fokker-Planck equation, Eq.~\eqref{eq:FP}, of the form $\rs(x,D)\propto\ee^{-\alpha V(x,D)}=\ee^{-\alpha ky^2/2}$, where $\alpha$ is a constant to be determined. Since $\rs$ depends on $(x,D)$ only through $y$, we have
\begin{equation}
  \partial_x\rs = -\alpha ky\,\rs \quad,\quad \partial_D\rs = \alpha ky\,\rs \quad,\quad \partial_x^2\rs = \partial_D^2\rs = -\alpha k\bigl(1-\alpha ky^2\bigr)\rs .
\end{equation}
Substituting these into the Fokker-Planck equation yields
\begin{align}
  \partial_t\rs &= \frac{k}{\gamma}\partial_x\bigl(y\rs\bigr) - \frac{1}{\tau_f}\partial_D\bigl(y\rs\bigr) + \Dx\partial_x^2\rs + \DD\partial_D^2\rs \nonumber\\
  &= \Bigl[\frac{k}{\gamma}+\frac{1}{\tau_f}-\alpha k\bigl(\Dx+\DD\bigr)\Bigr]\bigl(1-\alpha ky^2\bigr)\rs ,
\end{align}
which vanishes for all $(x,D)$ if and only if
\begin{equation}
  \alpha = \frac{k/\gamma+1/\tau_f}{k(\Dx+\DD)}.
\end{equation}
Substituting Eq.~\eqref{eq:diffusion} into the above expression yields
\begin{equation}
    \alpha = \frac{2\beta\tau_f(\gamma+k\tau_f)}{k(\beta\gamma\sigma^2+2\tau_f^2)} = \beta^*.\label{app:eq:betastar}
\end{equation}
Therefore, the stationary solution of the Fokker-Planck equation is given by $\rs(x,D)\propto\ee^{-\beta^* V(x,D)}$, verifying Eq.~\eqref{eq:betastar}. The same result follows from the dynamics of the relative coordinate $y$. Subtracting the two equations in Eqs.~\eqref{app:eq:EOMharm}, the equation of motion for $y$ is
\begin{equation}
  \dot y = -\Bigl(\frac{k}{\gamma}+\frac{1}{\tau_f}\Bigr)y + \sqrt{2\Dx}\,\xi_x(t) - \sqrt{2\DD}\,\xi_D(t) ,
  \label{app:eq:EOM-y}
\end{equation}
which recovers Eq.~\eqref{eq:wy} for $y$. Since $\xi_x$ and $\xi_D$ are independent Gaussian white noises, the combined noise term is equivalent to a single Gaussian white noise with intensity $2(\Dx+\DD)$. Hence, $y$ is an Ornstein-Uhlenbeck process, and its probability density $\rho(y,t)$ evolves according to the Fokker-Planck equation
\begin{equation}
  \partial_t\rho(y,t) = \Bigl(\frac{k}{\gamma}+\frac{1}{\tau_f}\Bigr)\partial_y\bigl(y\,\rho(y,t)\bigr) + \bigl(\Dx+\DD\bigr)\partial_y^2\rho(y,t) .
  \label{app:eq:FP-y}
\end{equation}
Direct substitution into Eq.~\eqref{app:eq:FP-y} verifies that
\begin{equation}
  \rs(y) = \sqrt{\frac{\beta^*k}{2\pi}}\,\ee^{-\beta^*ky^2/2}
\end{equation}
is the stationary distribution of $y$, and the corresponding stationary mean and variance are
\begin{equation}
    \langle y\rangle^s=0 \quad, \quad\langle y^2\rangle^s=1/(\beta^*k), \label{app:eq:stat-y}
\end{equation}
which are also given in Eq.~\eqref{eq:stat-y}.

Substituting $\rs$ into Eq.~\eqref{eq:currents} yields the stationary currents
\begin{equation}
  \Jx = \Bigl(\beta^*k\Dx-\frac{k}{\gamma}\Bigr)y\,\rs \quad,\quad \JD = \Bigl(\frac{1}{\tau_f}-\beta^*k\DD\Bigr)y\,\rs .
\end{equation}
By Eqs.~\eqref{app:eq:betastar} and \eqref{eq:diffusion}, the two prefactors are equal, and we have
\begin{equation}
  \Jx = \JD = \frac{k}{\gamma\beta}\bigl(\beta^*-\beta\bigr)\,y\,\rs .
  \label{app:eq:harm-currents}
\end{equation}
The stationary currents vanish if and only if $\beta^*=\beta$. Using $\partial_x(y\rs)=-\partial_D(y\rs)=(1-\beta^*ky^2)\rs$, we arrive at
\begin{equation}
  \partial_x\ln\rs = -\partial_D\ln\rs = -\beta^*ky \quad,\quad
  \frac{\partial_x\Jx}{\rs} = -\frac{\partial_D\JD}{\rs} = \frac{k}{\gamma\beta}\bigl(\beta^*-\beta\bigr)\bigl(1-\beta^*ky^2\bigr) , \label{app:eq:harm-derivs-J}
\end{equation}
which are reported in Eqs.~\eqref{eq:harm-derivs} of the main text.

\subsection{Heat and work}
\label{app:harm-QW}

The heat and work rates are defined in Eq.~\eqref{eq:QW-def} as Stratonovich products, for which the ordinary chain rule gives the first law of thermodynamics,
\begin{equation}
    \dot V=\dot x\strat\partial_xV+\dot D\strat\partial_DV=\dot Q+\dot W.
\end{equation}
To obtain the explicit expressions of $\dot Q$ and $\dot W$ in the It\^o convention, we apply It\^o's lemma to $V(x,D)$,
\begin{equation}
  \dd V = \partial_xV\,\dd x + \Dx\partial_x^2V\,\dd t + \partial_DV\,\dd D + \DD\partial_D^2V\,\dd t ,
\end{equation}
where the mixed term $\partial_x\partial_DV\,\dd x\,\dd D$ vanishes as $\xi_x$ and $\xi_D$ are independent. Therefore, the rate of change of $V$ is given by
\begin{equation}
  \dot V = \dot x\,\partial_xV + \Dx\partial_x^2V + \dot D\,\partial_DV + \DD\partial_D^2V ,
\end{equation}
where the products are now understood in the It\^o convention. The contributions associated with the dynamics of $x$ and of $D$ are the heat and the work, respectively,
\begin{equation}
  \dot Q = \dot x\,\partial_xV + \Dx\partial_x^2V \quad,\quad \dot W = \dot D\,\partial_DV + \DD\partial_D^2V .
  \label{app:eq:QW-ito}
\end{equation}
Substituting $V= k(x-D)^2/2$, and the equations of motion, Eqs.~\eqref{app:eq:EOMharm}, together with Eq.~\eqref{eq:diffusion}, we obtain
\begin{subequations}
\label{app:eq:QW}
\begin{align}
  \dot Q &= -\frac{k^2}{\gamma}y^2 + \frac{kT}{\gamma} + ky\sqrt{\frac{2T}{\gamma}}\,\xi_x , \\
  \dot W &= -\frac{k}{\tau_f}y^2 + \frac{k\sigma^2}{2\tau_f^2} - \frac{k\sigma}{\tau_f}\,y\,\xi_D ,
\end{align}
\end{subequations}
which are Eqs.~\eqref{eq:QW} of the main text.

\subsection{Stationary means}
\label{app:harm-means}

In the stationary state, the mean of $\dot Q$ is evaluated as
\begin{equation}
  \langle \dot Q\rangle^s = -\frac{k^2}{\gamma}\langle y^2\rangle^s + \frac{kT}{\gamma} + k\sqrt{\frac{2T}{\gamma}}\,\langle y\,\xi_x\rangle^s .
\end{equation}
In the It\^o convention, the relative coordinate $y$ is independent of the noises at the same time, hence the last term vanishes. Using Eq.~\eqref{app:eq:stat-y}, together with Eq.~\eqref{app:eq:betastar}, we obtain
\begin{equation}
  \langle \dot Q\rangle^s = -\frac{k^2}{\gamma}\,\frac{\gamma\sigma^2+2T\tau_f^2}{2\tau_f(\gamma+k\tau_f)} + \frac{kT}{\gamma} = -\frac{k(k\sigma^2-2T\tau_f)}{2\tau_f(\gamma+k\tau_f)} . \label{app:eq:mean-dot-Q}
\end{equation}
Following the same procedure, with $\langle y\,\xi_D\rangle^s=0$, the stationary mean of $\dot W$ is
\begin{equation}
  \langle \dot W\rangle^s = -\frac{k}{\tau_f}\langle y^2\rangle^s + \frac{k\sigma^2}{2\tau_f^2} = \frac{k(k\sigma^2-2T\tau_f)}{2\tau_f(\gamma+k\tau_f)} .
\end{equation}
These are reported in Eq.~\eqref{eq:Qmean} of the main text. Note that $\langle \dot Q\rangle^s+\langle \dot W\rangle^s=0$, consistent with the first law in the stationary state.

\section{Harmonic feedback model: entropy production and information flow}
\label{app:harm-entropy}

In this appendix, we derive the subsystem entropy production rates of the harmonic feedback model, Eqs.~\eqref{eq:rates-harm}, their stationary means, Eqs.~\eqref{eq:rates-mean} and \eqref{eq:Stotmean}, and the stationary information flows, Eq.~\eqref{eq:Iflow-harm}. For convenience, we define
\begin{equation}
  \beta_D \equiv \frac{2\tau_f}{k\sigma^2} = \frac{1}{T_D} ,
  \label{app:eq:betaD}
\end{equation}
which is consistent with Eq.~\eqref{eq:TD} of the main text. 

\subsection{Subsystem entropy production rates}
\label{app:harm-rates}

We begin with computing $\Delta\dot S_x$. Substituting $F/T=-\beta ky$, where $\beta=1/T$, and Eq.~\eqref{app:eq:harm-derivs-J} into Eq.~\eqref{eq:Sxpm}, the rate of $\Delta S_x$ becomes
\begin{equation}
  \Delta\dot S_x = (\beta^*-\beta)\,k\,\dot x\strat y + \frac{k}{\gamma\beta}\bigl(\beta^*-\beta\bigr)\bigl(1-\beta^*ky^2\bigr) .
\end{equation}
Since $k\,\dot x\strat y=\dot Q$ by Eq.~\eqref{eq:QW-def}, we obtain
\begin{equation}
  \Delta\dot S_x = (\beta^*-\beta)\Bigl[\dot Q + \frac{k}{\gamma\beta}\bigl(1-\beta^*ky^2\bigr)\Bigr] .
  \label{app:eq:Sxdot}
\end{equation}
By the same procedure, substituting $2\tau_f(O-D)/\sigma^2=\beta_Dky$ and Eq.~\eqref{app:eq:harm-derivs-J} into Eq.~\eqref{eq:SDpm}, the rate of $\Delta S_D$ becomes
\begin{equation}
  \Delta\dot S_D = -(\beta^*-\beta_D)\,k\,\dot D\strat y - \frac{k}{\gamma\beta}\bigl(\beta^*-\beta\bigr)\bigl(1-\beta^*ky^2\bigr) .
\end{equation}
Since $-k\,\dot D\strat y=\dot W$ by Eq.~\eqref{eq:QW-def}, we have
\begin{equation}
  \Delta\dot S_D = (\beta^*-\beta_D)\,\dot W - \frac{k}{\gamma\beta}\bigl(\beta^*-\beta\bigr)\bigl(1-\beta^*ky^2\bigr) .
  \label{app:eq:SD-inter}
\end{equation}
Using Eqs.~\eqref{app:eq:betastar} and \eqref{app:eq:betaD}, one can verify that
\begin{equation}
  -\frac{k}{\gamma\beta}\bigl(\beta^*-\beta\bigr) = \frac{k\sigma^2}{2\tau_f^2}\bigl(\beta^*-\beta_D\bigr) .
  \label{app:eq:prefactor}
\end{equation}
Substituting this into Eq.~\eqref{app:eq:SD-inter} yields
\begin{equation}
  \Delta\dot S_D = (\beta^*-\beta_D)\Bigl[\dot W + \frac{k\sigma^2}{2\tau_f^2}\bigl(1-\beta^*ky^2\bigr)\Bigr] .
  \label{app:eq:SDdot}
\end{equation}
Equations~\eqref{app:eq:Sxdot} and \eqref{app:eq:SDdot} are Eqs.~\eqref{eq:rates-harm} of the main text. Finally, substituting the explicit expressions of $\dot Q$ and $\dot W$, Eqs.~\eqref{app:eq:QW}, we obtain the explicit expressions of $\Delta\dot S_x$ and $\Delta\dot S_D$ in the It\^o convention,
\begin{subequations}
\label{app:eq:rates-ito}
\begin{align}
  \Delta\dot S_x &= (\beta^*-\beta)\frac{k}{\gamma}\Bigl[2T-\bigl(1+\beta^*T\bigr)ky^2+\sqrt{2\gamma T}\,y\,\xi_x\Bigr] , \\
  \Delta\dot S_D &= (\beta^*-\beta_D)\,k\Bigl[\DD\bigl(2-\beta^*ky^2\bigr)-\frac{y^2}{\tau_f}-\frac{\sigma}{\tau_f}\,y\,\xi_D\Bigr] ,
\end{align}
\end{subequations}
which depend only on $y$ and the noises.

\subsection{Stationary means}
\label{app:harm-rate-means}

The terms proportional to $1-\beta^*ky^2$ in Eqs.~\eqref{app:eq:Sxdot} and \eqref{app:eq:SDdot} have zero stationary mean, since $k\langle y^2\rangle^s=1/\beta^*$ by Eq.~\eqref{app:eq:stat-y}. Together with $\langle \dot W\rangle^s=-\langle \dot Q\rangle^s$, the stationary means are
\begin{equation}
  \langle \Delta\dot S_x\rangle^s = (\beta^*-\beta)\langle \dot Q\rangle^s \quad,\quad
  \langle \Delta\dot S_D\rangle^s = (\beta_D-\beta^*)\langle \dot Q\rangle^s \quad,\quad
  \langle \Delta\dot S_{\mathrm{tot}}\rangle^s = (\beta_D-\beta)\langle \dot Q\rangle^s .
  \label{app:eq:rate-means}
\end{equation}
To evaluate them explicitly, we use Eqs.~\eqref{app:eq:betastar} and \eqref{app:eq:betaD} to compute
\begin{equation}
  \beta^*-\beta = \frac{\beta\gamma(2\tau_f-\beta k\sigma^2)}{k(\beta\gamma\sigma^2+2\tau_f^2)} \quad,\quad
  \beta_D-\beta^* = \frac{2\tau_f^2(2\tau_f-\beta k\sigma^2)}{k\sigma^2(\beta\gamma\sigma^2+2\tau_f^2)} \quad,\quad
  \beta_D-\beta = \frac{2\tau_f-\beta k\sigma^2}{k\sigma^2} .
\end{equation}
Substituting these, together with Eq.~\eqref{app:eq:mean-dot-Q}, into Eq.~\eqref{app:eq:rate-means}, we obtain
\begin{subequations}
\begin{align}
  \langle \Delta\dot S_x\rangle^s &= \frac{\gamma(\beta k\sigma^2-2\tau_f)^2}{2\tau_f(\beta\gamma\sigma^2+2\tau_f^2)(\gamma+k\tau_f)} , \\
  \langle \Delta\dot S_D\rangle^s &= \frac{\tau_f(\beta k\sigma^2-2\tau_f)^2}{\beta\sigma^2(\beta\gamma\sigma^2+2\tau_f^2)(\gamma+k\tau_f)} , \\
  \langle \Delta\dot S_{\mathrm{tot}}\rangle^s &= \frac{(\beta k\sigma^2-2\tau_f)^2}{2\beta\tau_f\sigma^2(\gamma+k\tau_f)} ,
\end{align}
\end{subequations}
which are reported in Eqs.~\eqref{eq:rates-mean} and \eqref{eq:Stotmean} of the main text. All three means are nonnegative and vanish if and only if $\beta k\sigma^2=2\tau_f$, i.e., $T=T_D$.

\subsection{Information flow}
\label{app:harm-info}

For the present model, Eqs.~\eqref{eq:Smed} and \eqref{eq:Smf} give
\begin{equation}
  \Delta\dot S_{\mathrm{med}} = -\beta k\,\dot x\strat y = -\beta\dot Q \quad,\quad
  \Delta\dot S_{\mathrm{m\&f}} = \beta_Dk\,\dot D\strat y = -\beta_D\dot W ,
  \label{app:eq:Smed-Smf-harm}
\end{equation}
which are the same as Eqs.~\eqref{eq:Smed-Smf-harm} of the main text. Substituting these and Eqs.~\eqref{app:eq:rate-means} into Eqs.~\eqref{eq:Iflow-S}, together with $\langle \dot W\rangle^s=-\langle \dot Q\rangle^s$, yields
\begin{align}
  \langle \dot I^s_x\rangle^s &= -\beta\langle \dot Q\rangle^s - (\beta^*-\beta)\langle \dot Q\rangle^s = -\beta^*\langle \dot Q\rangle^s , \nonumber\\
  \langle \dot I^s_D\rangle^s &= \beta_D\langle \dot Q\rangle^s - (\beta_D-\beta^*)\langle \dot Q\rangle^s = \beta^*\langle \dot Q\rangle^s ,
\end{align}
which is Eq.~\eqref{eq:Iflow-harm} of the main text.

\twocolumngrid

\bibliography{refs}

\end{document}